# A Scalable and Adaptive Method for Finding Semantically Equivalent Cue Words of Uncertainty

Chaomei Chen[a,b], Ming Song[*,b], Go Eun Heo[b]
[a]College of Computing and Informatics, Drexel University, USA
[b]Department of Library and Information Science, Yonsei University, South Korea

## Abstract
Scientific knowledge is constantly subject to a variety of changes due to new discoveries, alternative interpretations, and fresh perspectives. Understanding uncertainties associated with various stages of scientific inquiries is an integral part of scientists' domain expertise and it serves as the core of their meta-knowledge of science. Despite the growing interest in areas such as computational linguistics, systematically characterizing and tracking the epistemic status of scientific claims and their evolution in scientific disciplines remains a challenge. We present a unifying framework for the study of uncertainties explicitly and implicitly conveyed in scientific publications. The framework aims to accommodate a wide range of uncertain types, from speculations to inconsistencies and controversies. We introduce a scalable and adaptive method to recognize semantically equivalent cues of uncertainty across different fields of research and accommodate individual analysts' unique perspectives. We demonstrate how the new method can be used to expand a small seed list of uncertainty cue words and how the validity of the expanded candidate cue words are verified. We visualize the mixture of the original and expanded uncertainty cue words to reveal the diversity of expressions of uncertainty. These cue words offer a novel resource for the study of uncertainty in scientific assertions.

## Keywords
Uncertainty, semantically equivalent words, scientific assertions, deep learning, resources

## Highlights

- A generalized framework of uncertainty is presented to accommodate uncertainties due to inconsistent and contradictory information as well as those associated with hedging and other linguistically focused cues.
- A scalable and adaptive method selects semantically equivalent words.
- The method advances the selection of scientific assertions involving uncertainties.
- The study offers new resources for studying the role of uncertainty in science.

## Introduction
A scientific proposition is a statement such as smoking causes cancer. The epistemic status of a scientific proposition refers to the best knowledge of its truthfulness given the current scientific knowledge. Thus, the epistemic status may range from completely unknown to speculations and from hypotheses to facts. The concept of uncertainty in this context characterizes the lack of sufficient information on a given proposition. A statement concerning a proposition can be considered as a combination of two parts: the proposition proper and information relevant to the epistemic status of the proposition. In this article, we focus on uncertainties due to lack of information and, in particular, uncertainties due to lack of consensus. Scientists routinely deal with such uncertainties at various stages of their research, from formulating research questions and selecting research methods to interpreting their findings and communicating their work to others (Cordner & Brown, 2013). Light et al. (2004) estimated that 11% of sentences in MEDLINE abstracts are speculative. Sociologists have studied the formation of consensus in the scientific community concerning whether smoking indeed causes cancer and whether a consensus is reached on climate change (Shwed & Bearman, 2010). Scientists face intensified uncertainties when inconsistent, conflicting, or contradictory findings emerge and when competing paradigms are proposed to resolve pressing crises (Kuhn, 1970). The formation of a consensus or the establishment of a dominant paradigm may correspond to a decrease of the overall uncertainty associated with a field of research. However, as we all know, searching for answers to seemingly simple questions may quickly lead to many much more complicated questions. The ability to assess the state of the art of a field of research effectively and efficiently at various levels of granularity is crucial for scientists, science policy makers, and the public.

Research in computational linguistics has made significant advances in identifying uncertainty cues and negations. Remarkably influential efforts include the development of the BioScope Corpus for uncertainty and negation in biomedical publications (Vincze et al., 2008), the CoNLL 2010 Shared Task (Farkas et al., 2010) for detecting hedges and their scope in natural language text, the enrichment of a biomedical event corpus with meta-knowledge (Thompson et al., 2011), and unifying categorizations of semantic uncertainty for cross-genre and cross domain uncertainty detection (Szarvas et al., 2012).

For example, the CoNLL-2010 shared task (Farkas et al., 2010) focused on detection of uncertainty cues and its linguistic scope in natural language texts. Typical hedging cue is composed of four categories: 1) auxiliaries, 2) verbs of hedging or verbs with speculative content, 3) adjectives or adverbs, and 4) conjunctions. Uncertainty detection focused on biomedical articles and text on Wikipedia. The best uncertainty detection performance in the CoNLL-2010 shared task was achieved with sequence labeling (e.g., Conditional Random Fields) in the biomedical data and bag of words sentence classification in the Wikipedia data. For the in-sentence hedge scope detection task, they classify each token to detect specific cue scopes. Their system is different from the number of class label used target and machine learning approach. More recent studies have explored the potential of measuring the confidence of biomedical models such as pathways based on textual uncertainty (Zerva et al., 2017) and the feasibility of assessing the factuality of predications extracted by SemRep (Kilicoglu et al., 2017).

In a broader context, identifying and measuring the degree of uncertainties associated with scientific knowledge embedded in the vast and fast-growing volume of scientific literature remain a bottleneck (Chen, 2016). Influential computational linguistic approaches such as hedging (Hyland, 1998), semantic uncertainty (Szarvas et al., 2012), negation (Chapman et al., 2001; Morante & Daelemans, 2009), and discourse-level uncertainty (Vincze, 2013) have been largely motivated by issues concerning uncertainties from linguistic perspectives. As demonstrated by Simmerling (2015), by using grammatical, stylistic, and rhetorical options, one can talk about scientific uncertainty without using any lexical cues of uncertainty. Furthermore, philosophical and sociological studies of science, scientific creativity, and scientific discovery have highlighted the role of identifying and resolving contradictions and inconsistencies in scientific discovery and in divergent thinking in general. In particular, the value of reconciling multiple perspectives has been long recognized and advocated (Collins, 1989; Linstone, 1981). It is critical for scientists to be able to track conflicting views on the same issue and resolve seemingly contradictory evidence at a new level (Chen, 2014, 2016). The linguistically motivated approaches to the study of scientific uncertainty may benefit from a broadened scope of perspectives.

In this article, we present a conceptual framework of the study of uncertainty based on a novel conceptualization of uncertainty as an epistemic status of scientific propositions. The new conceptualization underlines the nature of uncertainty as a meta-knowledge of science and its integral role in scientific change. We introduce a scalable and adaptive method to identify uncertainty cues under the broadened conceptualization of uncertainty. The resultant uncertainty cue words are expected to provide a useful resource for further studies of scientific uncertainty. The method is adaptive in the sense that analysts may generate semantically equivalent uncertainty cues of new dimensions based on a small number of example words.

The rest of the article is organized as follows. First, we introduce basic concepts concerning scientific propositions and illustrate some of the most common types of uncertainties associated semantic predications in MEDLINE and the distributions of leading uncertainty cue words in other collections of scientific publications. Next, we present a scalable and adaptive method to construct a comprehensive set of uncertainty cue words from scientific publications. The method begins with a set of hand-crafted uncertainty cue words as seeds based on a general-purpose thesaurus of English. Then the computational method expands the seed list to a much larger set of semantically equivalent uncertainty cue words. Two judges evaluated the expanded cue words. The accepted and rejected cue words along with the seed words are visualized as non-overlapping clusters. Sample sentences selected by these uncertainty cues are discussed. The collection of the specific uncertainty cue words, classes of these words, and corresponding statistics are provided as a community resource for researchers to build on the result of our research.

## Uncertainties of Scientific Knowledge

Scientific knowledge is a complex adaptive system of facts, beliefs, hypotheses, speculations, opinions, and a wide variety of other types of information about what we know and how much we know. It is adaptive in that existing scientific knowledge is subject to re-examination in light of new discoveries, alternative interpretations, and scenarios that are previously thought impossible (Chen, 2014; Popper, 1961). A scientist's domain expertise consists of not only his or her knowledge of various facts and consensus in science but also an accurate understanding of the epistemic status of a wide variety of unsettled elements of a scientific domain. The epistemic status of a scientific proposition characterizes various stages of its epistemological advances driven by underlying scientific inquiries. For example, our beliefs of the truthfulness of a proposition may vary significantly based on available evidence in scientific literature, ranging from anecdotes and case studies of a small sample to the support of large-scale meta-analyses of randomized double blind clinical trials. Scientists often need to deal with conflicting and contradictory findings concerning the same propositions (Ioannidis & Trikalinos, 2005). Dealing with scientific uncertainties is the norm in the development of science rather than the exception.

Major sources of scientific uncertainty at macroscopic levels have been studied across a diverse range of disciplines. Sociological theories of scientific change, for example, underline the tension between the novelty of a research topic and its potential for scientists to compete for their reputations (Fuchs, 1993). Potentially highly rewarding research tends to have high uncertainties and high risks. Early stages of an emerging field of research tend to involve a high level of uncertainty (Shneider, 2009). The uncertainty level of a scientific field is particularly high when it is experiencing fundamental crises, which may trigger a scientific revolution or a paradigm shift (Kuhn, 1970). Shwed and Bearman (2010) show that the formation of scientific consensus may follow as scientific rivalries consider a proposition to be a fact.

### *Hedging*

Hedging is a particularly relevant concept in understanding how scientists characterize the tentative and context-dependent nature of scientific claims (Lakoff 1973; Vold, 2006). The use of hedge words has been intensively studied in terms of their role as uncertainty cues (e.g., Hyland, 1998). Hedging is considered as a sign of uncertainty that authors would like to attribute to their assertions. Commonly used hedging words include *may*, *could*, *might*, as well as other words such as *suggest*, *indicate*, *appear*, *seem*, and *assume*. Hedge words include adjectives, nouns, verbs, and modal verbs. Hedging can mitigate an otherwise overstated scientific claim such that the epistemic status of speculations and facts can be communicated clearly.

Citing the original source is considered a type of hedging because the burden is shifted to the author of the original source. Horn (2001) revealed that when scientists paraphrase assertions containing hedges from publications in the literature, they often omit the original hedges. Such omissions may distort the uncertainty expressed in the original assertions.

Algorithmically identifying hedging and negation in scientific publications, especially in biomedical domains, has been extensively investigated by a series of influential studies over the last ten years. For example, the BioScope corpus (Vincze et al., 2008) has been instrumental for the development of computational linguistic tools to detect uncertainty and negation cues and their scopes in biomedical documents. The CoNLL 2010 Shared Task for detecting hedges as uncertainty cues in natural language texts (Farkas et al., 2010) has generated a long-lasting impact, for example, leading to the study of weasel words (Vincze, 2013), detecting negation and speculation for sentiment analysis (Cruz et al., 2016), assessing factuality drift in resolved rumors on Twitter (Lendvia et al., 2016), argumentation mining (Habernal & Gurevych, 2017) , and assessing the confidence in biomedical pathways (Zerva et al., 2017).

Computational linguistic approaches to the detection of uncertainty, negation, and speculation cues include patterns specified by hand-crafted rules, supervised learning and semi-supervised learning techniques, and multi-level classifiers (Szarvas et al., 2012; Malhotra et al., 2013). Currently, the majority of the study of uncertainty in scientific articles is linguistically motivated (e.g. Kilicoglu & Bergler, 2008).

Thompson et al. (2011) enriched a biomedical event corpus with an annotation scheme of meta-knowledge. A biomedical event refers to "representations of important facts and findings contained within documents" and the relevant meta-knowledge refers to information that can be derived from the context of the event. Their meta-knowledge annotation scheme contains several dimensions of meta-knowledge, include three certainty levels (speculation, probable, certain), two levels of polarity (positive or negative), and six types of knowledge (investigation, observation, analysis, method, fact, and other).

Szarvas et al. (2012) categorized semantic uncertainties into two major categories of epistemic and hypothetical uncertainties. Hypothetical uncertainties are in turn divided into paradoxical and non-epistemic modality. Paradoxical uncertainties contain investigation and condition as sub-categories. Under the non-epistemic modality, there are doxastic and dynamic uncertainties. Szarvas et al. (2012) normalized the annotation of three corpora for recognizing uncertainty cues across genres and domains.

Most of the computational linguistic studies we have reviewed do not explicitly single out propositions and meta-knowledge from natural language texts. A notable exception is Semantic MEDLINE, which includes explicit representations of propositions extracted from MEDLINE abstracts (Kilicoglu, Shin, Fiszman, Rosemblat, & Rindflesch, 2012; Rindflesch & Fiszman, 2003). In Semantic MEDLINE, propositions are known as semantic predications.

### *Uncertainties of Semantic Predications*

Semantic MEDLINE is a repository of semantic predications extracted by SemRep from MEDLINE titles and abstracts. A semantic predication in Semantic MEDLINE is a subject-predicate-object triple. For example, "HIV CAUSES AIDS" is a semantic predication. The subject HIV and the object AIDS are UMLS concepts. Each UMLS concept represents a group of instances of the same underlying concept. For example, HIV as a concept represents a group of instances of the concept in natural language, including human immunodeficiency virus, lymphadenopathy-associated virus, AIDS virus, HIV-1LAI, and HTLVIII. Predications are pre-defined semantic types such as CAUSES, AFFECTS, and PART_OF. The negation of a semantic predicate is represented by the prefix NEG_ for the predicate. For example, the negation of "HIV CAUSES AIDS" is "HIV NEG_CAUSES AIDS." Semantic MEDLINE provides a valuable repository of semantic predications and enables researchers to analyze scientific knowledge at multiple levels of granularity in areas such as literature-based discovery (Cameron et al., 2013) and drug-disease-gene patterns (Zhang et al., 2014).

Analyzing semantic predications in Semantic MEDLINE has also drawn our attention to the role of a scientific proposition and its epistemic status. The following sentence is from the abstract of a MEDLINE record. The sentence contains four propositions, which are asserted with different levels of uncertainty. The first two propositions are concerning the role of Helicobacter pylori gastritis in gastritis and duodenal ulcer. The two propositions are represented by two semantic predications in Semantic MEDLINE (highlighted in the sentence in boldface): 1) Helicobacter-associated gastritis AFFECTS Gastritis and 2) Helicobacter-associated gastritis AFFECTS Duodenal Ulcer. Like propositions, these semantic predications do not contain any hedging. Their truthfulness, however, is expressed in the original sentence. The phrase "the established role" qualifies the epistemic status of the two propositions – both of them are accepted facts. In contrast, uncertainties are indicated in the second half of the sentence.

> *In contrast to the established role of* **Helicobacter pylori gastritis** *in* **gastritis** *and* **duodenal ulcer** *in general, conflicting results have been reported in patients with human immunodeficiency virus (HIV) infection and the acquired immunodeficiency syndrome.*

The second half of the sentence is associated with two more predications (underlined in the sentence): 3) HIV Infections PROCESS_OF Patients and 4) Acquired Immunodeficiency Syndrome PROCCESS_OF Patients. This part of the sentence has several issues. The sentence does not specifically identify the source of the "conflicting results." This is an example of weasels because it omits the information on who reported the conflicting results and it does not cite any reference. More importantly, it is ambiguous about the specific proposition to which the conflicting results are attributed. Does it refer to the role of Helicobacter pylori gastritis in patients with AIDS? At least, it is difficult to resolve the ambiguity without a broader context. The uncertainty cues contained in the sentence, namely the transitional "in contrast" to an established fact, the passive tone about conflicting results, and the phrase "conflicting results" differ in terms of their strength as a signal for uncertainty. The strength of the phrase "conflicting results" is the strongest because it refers to the epistemic status of the proposition in question specifically and explicitly.

The uncertainty due to conflicting results meets the description of the category of epistemic uncertainty because "on the basis of our world knowledge we cannot decide at the moment whether it is true or false" (Szarvas et al. 2012). On the other hand, it seems to fit the more specific paradoxical sub-category of the hypothetical uncertainty category because the mixed signals of the truthfulness of the proposition in question and the epistemic status of the proposition is unsettled. Ambiguities in how one should categorize instances of uncertainty cues at multiple levels of granularity are likely to hinder the annotation of uncertainty cues.

Researchers have proposed various scales to organize lexical cues for uncertainty and speculation. For example, In HypothesisFinder, Malhotra et al. (2013) identified three groups of cues for speculation based on their efficacy in recognizing a speculative sentence. Strong patterns include "*might be involved*," "*hypothesized that*," and "*raising the possibility that*." Moderate patterns include "*seems to*," "*appears to be*," and "*can be anticipated*." Weak patterns include "*presume*," "*suppose*," and "*would*." Malhotra et al. (2013) found that combining with additional cue words of speculation or hedges can improve the performance of weak patterns, which are mostly single word and may lead to false positives due to lack of specificity. Thompson et al. (2011) defined three categories of uncertainty of an event, which plays a similar role as a proposition in this article. Each level is defined based on two criteria: either the extent of uncertainty or speculation or the frequency of the event in question: L3) no explicit indication of uncertainty or speculation or the frequency is high, L2) high confidence in terms of likelihood or the event occurs frequently, and L1) low confidence or the event occurs rarely.

The heterogeneity of available datasets annotated with uncertainty cues across subject domains is one of the challenges for the development of uncertainty detection applications in new domains (Szarvas et al., 2012). Szarvas et al. (2012) estimated that training an accurate uncertainty cue detector for a new domain or a new genre may require a manually annotated training data set of 3,000~5,000 sentences.

## A Conceptual Framework of Scientific Uncertainty

Motivated by relevant studies discussed above concerning various uncertainties found in scientific publications, especially the categorization of uncertainty (Szarvas et al., 2012) and meta-knowledge (Thompson et al., 2011), we propose a conceptual framework for the study of scientific uncertainty with an emphasis on uncertainties due to inconsistent and conflicting findings as the epistemic status of scientific knowledge.

Figure 1 illustrates the major components of the conceptual framework. Scientific knowledge consists of two types of information: A) propositions and B) meta-knowledge of propositions in terms of their epistemic status and perturbation strength. The scope of the meta-knowledge can be further expanded. The framework broadens the concept of uncertainty and characterizes it as an indicator of the epistemic status of a scientific proposition. The framework distinguishes scientific propositions and their epistemic status. The epistemic status of a proposition is the meta-knowledge of the proposition. The meta-knowledge may change with reduced uncertainties as we learn more about the truthfulness of a proposition.

*Epistemic Status of a Proposition*

The epistemic status of a proposition addresses questions concerning the truthfulness of a proposition. Is it true that smoking causes lung cancer? Is it trustworthy that MMR vaccine causes autism? Is there a consensus on how long Ebola virus may survive in water? The highest level of uncertainty is the complete unknown, whereas the lowest level of uncertain is associated with propositions that have been accepted as facts. In addition, different types of uncertainty differ in their potential to bring fundamental changes and revolutionize their field of research or to provide changes that are incremental in nature. For example, contradictions in scientific experiments may lead to breakthroughs and scientific revolutions. Therefore contradictions and inconsistencies are examples of the types of uncertainties that are strong in their perturbation strength when we conceptualize the scientific knowledge as a complex adaptive system.

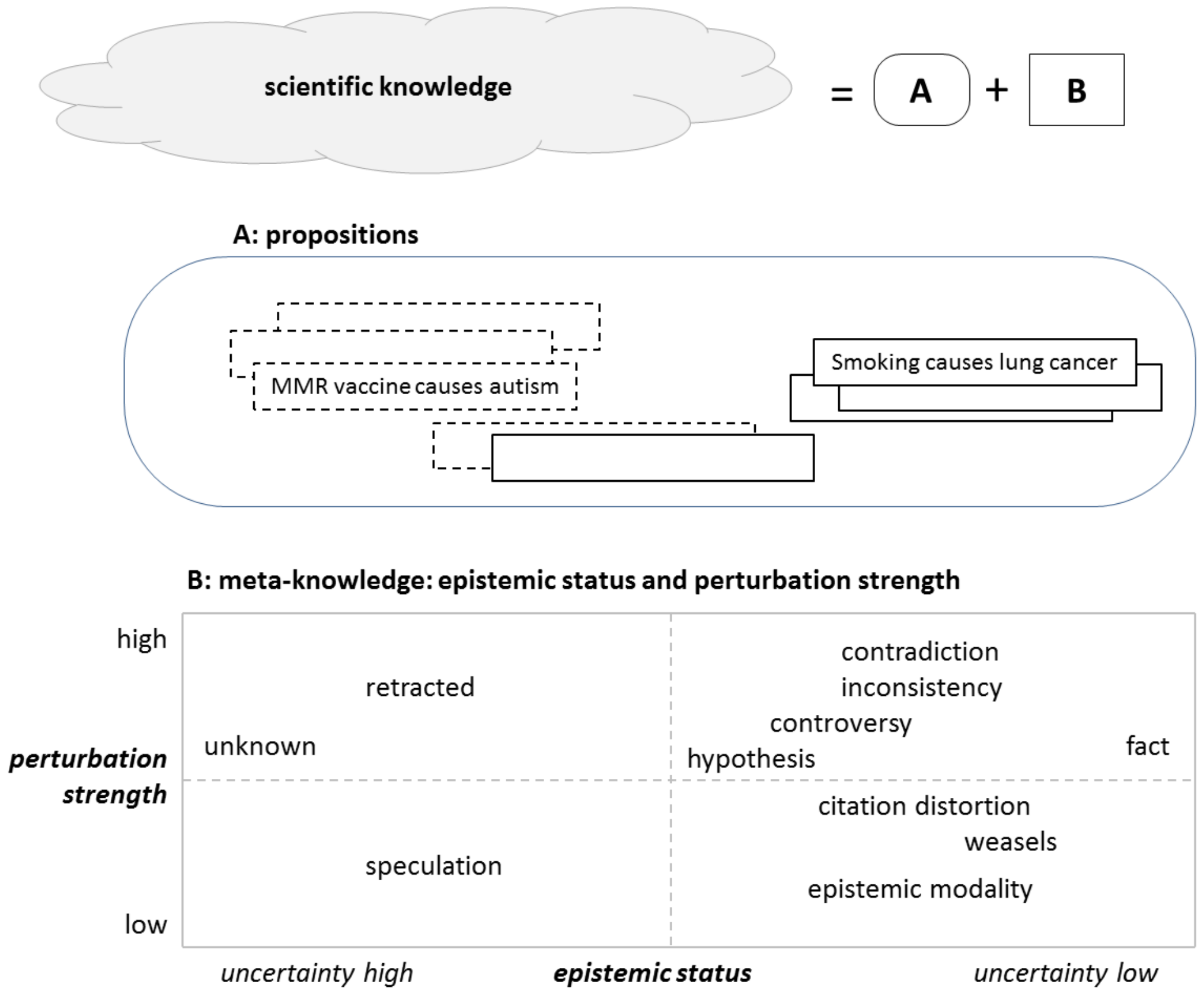

**Figure 1. An illustrative sketch of the conceptual framework for the study of uncertainty in scientific literature.**

The framework considers uncertainties at the system level by synthesizing uncertainties at the level of individual assertions made in scientific literature. A scientific proposition is considered uncertain if the truthfulness of the proposition is questionable or unsettled. At the local level, a proposition is uncertain if there are any indications of lack of information concerning the true value of the proposition, ranging from the complete unknown to speculations, hypotheses, and a wide variety of hedge words.

We use semantic predications in Semantic MEDLINE to illustrate propositions and examples of how various types of uncertainties arise in their original texts. Semantic predications in Semantic MEDLINE can be seen as propositions because they assert a proposition without any contextual information. Figure 2 demonstrates semantic predications such as Virus CAUSES Infection and what additional information about them may tell us. The proposition Virus CAUSES Infection itself does not provide any information about whether it is true, false, or somewhere in between to the best of the scientific knowledge at a given point of time. In contrast, we may learn a lot about its epistemic status by examining various assertions concerning the proposition throughout the years in MEDLINE records.

The epistemic status of a proposition is a function of newly published research findings over time. Figure 2 shows a timeline visualization of burst detection. Propositions are listed in the first column. The multi-color lines on the right visualize the "bursts" detected with these propositions. Burst detection aims to identify events that occur with much higher frequencies – hence bursts – than their other events (Kleinberg, 2002). In this case, the burst of a proposition over time means that there was a period of time the proposition was particularly popular in MEDLINE. The period of burst is depicted by the line segment in red. The light blue line depicts the period prior to the first appearance of the proposition in question, whereas the darker blue line depicts the period after the period of burst ended. The information on the burstness of a proposition may provide a useful timeframe for studying the evolution of the epistemic status of the proposition. For example, the end of a period of burst may serve as a useful reference point: if the attention to the proposition has decreased, it may be a sign that the uncertainty of the proposition is no longer considered high enough to retain its competitiveness as a research topic (e.g. Fuchs, 1993). Similarly, if a consensus has been reached on a once controversial proposition, scientists are likely to disperse and pursue new research topics elsewhere as in cases such as mass extinctions research (Chen, 2006).

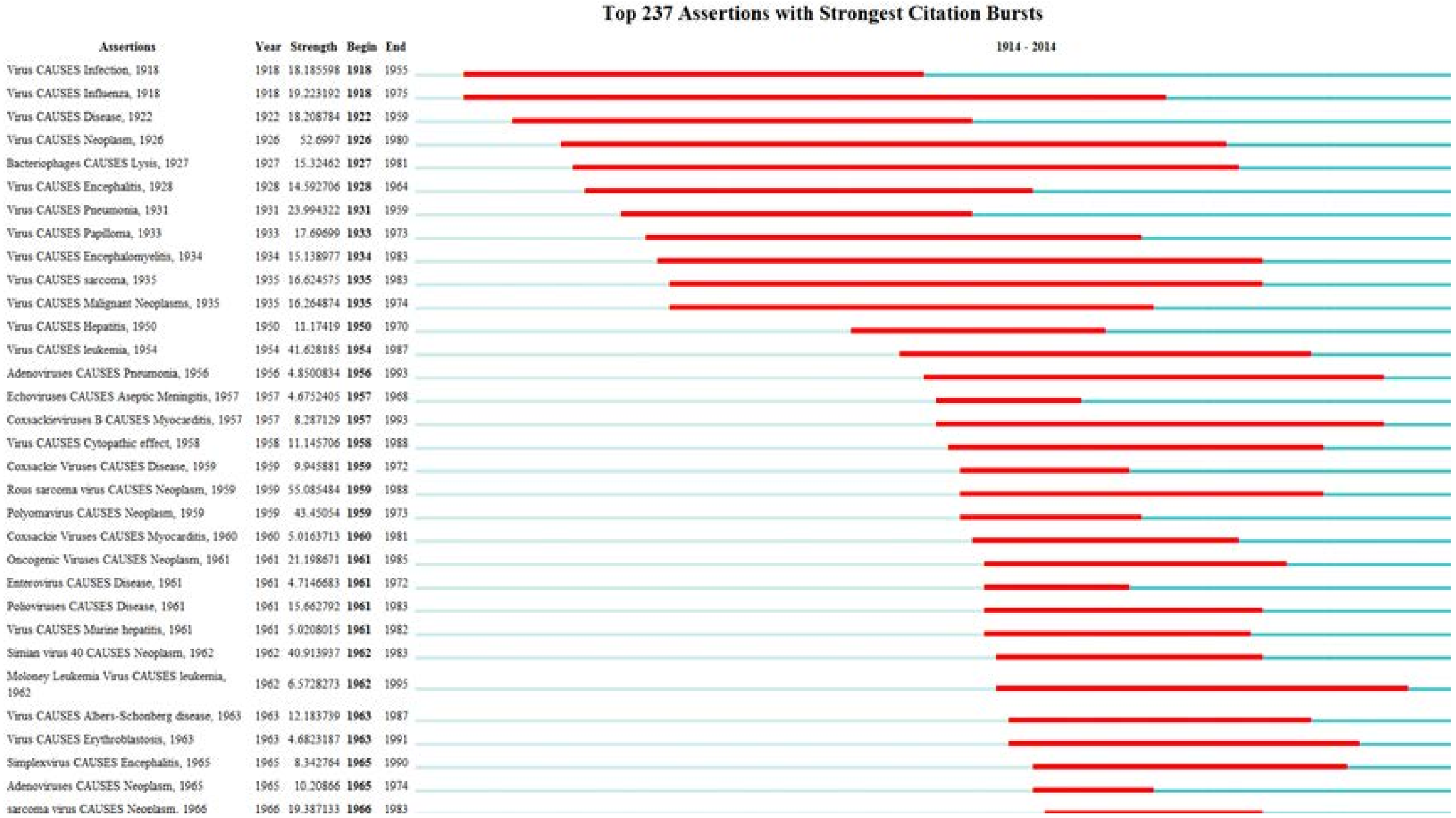

**Figure 2. Examples of propositions in scientific literature. The status of each of the propositions is time sensitive, which reflects our meta-knowledge of the underlying scientific knowledge.**

The temporal patterns of burst associated with propositions illustrate possibilities for integrating research on uncertainties of scientific propositions with at multiple levels of granularity.

### *Perturbation Strength*

In addition to uncertainty cues routinely studies in computational linguistic studies, indications of inconsistencies, contradictions, and controversies in the collective knowledge of the proposition in question are taken into account along the dimension of perturbation strength. Uncertainties due to inconsistencies, contradictions, and controversies are frequently studied under the subject of scientific uncertainties, especially in relation to decision making with uncertainties in topics such as global climate change (e.g. Zehr, 2000; Frewer et al., 2011). Scientific uncertainties play a more fundamental role in the development of science than uncertainties hinted by hedging according to philosophical and sociological theories of scientific change (e.g., Kuhn, 1970; Fuchs, 1993; Collins, 1989). Scientific uncertainties such as contradiction-induced ones may lead to potentially revolutionary changes of a scientific domain such as a paradigm shift or the emergence of a new field. In contrast, uncertainties that can be resolved by incrementally increasing our knowledge are relatively less critical because one may still retain the existing paradigm after all. A system-level uncertainty is possible even if individual assertions are made without any indications of uncertainty, for example, beliefs from researchers who belong to distinct schools of thought.

As a special case, the unknown is a valid value of the epistemic status. For example, the uncertainty conveyed by the sentence “The mechanism is unknown” is the highest. In general, one would expect the level of uncertainty associated with a proposition will be reduced as scientists investigate the topic further. However, there are several scenarios that may increase the level of uncertain regarding a proposition at the system level, i.e. to the collective knowledge of the scientific community as opposed to the uncertainty conveyed by a particular assertion. For example, when Kuhnian crises arise in a field of research, some of the fundamental propositions are challenged, which would lead to an increased level of uncertainty. Furthermore, when scientific publications are retracted, uncertainties of claims made in retracted articles would increase, especially before a scientific consensus is reached (Chen et al., 2013). For example, the partial retraction and the subsequent complete retraction of the article by Wakefield et al. (1998) may alter the uncertainty of the proposition “MMR vaccine CAUSES autism” because the retraction makes the proposition more doubtful. To a lesser extent, the epistemic status of a proposition may be distorted with false positives if original assertions’ hedges are dropped in later references (Horn, 2001).

## A Scalable and Adaptive Method for Generating Uncertainty Cues

The conceptual framework distinguishes propositions and their epistemic status at a particular time. Detecting a wide variety of uncertainty cues in natural language texts is a critical step for subsequent research and applications concerning scientific uncertainty, argumentation mining, and the dynamics of scientific knowledge. Computational linguistic studies often share strategies that start with manually identified uncertainty cues and then train classification models to identify additional cues and classify sentences of various uncertainties. As we have reviewed earlier, profound impacts have been made by computational linguistic studies, notably Chapman et al. (2001), Vincze et al. (2008), Thompson et al. (2011), Szarvas et al. (2012).

In this article, we introduce a scalable and adaptive method for finding semantically equivalent uncertainty cue words. The new method is motivated by several reasons. First, we need to incorporate scientific uncertainties that are caused by inconsistencies, contradictions, controversies, and other types of discrepancies in scientific literature. According to our conceptual framework, such uncertainties may lead to fundamental and revolutionary changes to scientific knowledge. Uncertainty cues used in existing studies do not adequately cover uncertainty cues specifically concerning inconsistencies and contradictions despite relevant categorizes such as paradoxical uncertainties have been identified (Thompson et al., 2011; Szarvas et al., 2012). Secondly, we are interested in an adaptive method in that it can be applied to distinct subject domains of interest with a minimal cost of manually annotating sentences for each subject domain. According to Szarvas et al. (2012), training computational linguistic models for a new domain may require 3,000~5,000 annotated sentences. The cost of manually annotating sentences in a highly technical domain by relatively inexperienced coders may be even higher due to the various ambiguities in natural language expressions, especially when uncertainties are involved. Another reason why manually annotating scientific uncertainties can be a serious challenge is the cognitive burden on the analyst to come up with as many expressions as possible so that machine learning algorithms can optimize their performance. The more examples we can feed to the algorithms, the better. However, listing a comprehensive list of possible uncertainty cues is an unrealistic task for a human analyst because of the complexity, contextual dependency, ambiguity, and diversity. After all, given the volume of scientific publications today, it is unrealistic to expect an individual to come up with a list that can cover a subject domain comprehensively.

We propose a method that starts with a small number of representative words as uncertainty cues and then expands to a much larger set of semantically equivalent words by using word2vec models (Mikolov et al., 2013) trained on large-scale documents.

### *Uncertainties due to Inconsistencies and Contradictions*

Commonly used hedging words provide signs of uncertainty to the extent that they are generally applicable across scientific disciplines. On the other hand, hedging words alone do not provide specific reasons that characterize the source of uncertainty. In contrast, "contradictory results" provides a useful explanation of the type of uncertainty involved.

In order to investigate the distributions of cue words of scientific uncertainties associated with discrepancies such as inconsistencies, conflicting opinions, contradictions, and controversies in scientific literature, we first construct two sets of sentences S+ and S- from MEDLINE. The set S+ consists of sentences that contain signs of scientific uncertainties, namely indicative words of conflicting results and contradictory findings. In contrast, S- consists of sentences that are free from these indicative words of inconsistency. In addition to identify potential cue words of uncertainty that may differ between S+ and S-, we are also interested in whether common hedge words are used differently between the two sets of sentences. If the distributions of hedge words remain the same in the two sets, then it suggests hedge words alone are unlikely to reflect the differences between the two sets.

The S+ set contains 35,572 sentences extracted from 33,880 MEDLINE records on virus. The S- set contains 35,527 sentences from 5,896 articles. We used hedge words from Hyland (1996) as a sample of the common hedge words.

Table 1 shows two lists of words that may be used to convey a sense of uncertainty. The list on the left-half of the table is from Hyland (1996), containing commonly recognized hedge words such as may, suggest, and might. The list on the right-half of the table is our extension in attempt to capture uncertainties concerning the status of a scientific inquiry, including the explicit use of the word uncertain and some of the other words that are commonly found when one describes a scientific investigation, including inconclusive, inconsistent, and hypothesis. The goal at this step is not to generate a comprehensive list of cue words for the latter types of uncertainties. Rather, the goal for this particular comparison is to illustrate how often assertions in scientific publications are concerned about giving a faithfully accurate description of what we know. No exaggerations and no distortions.

For each word on the two lists, its S+/S- ratio is the percentage of sentences containing the word in S+ to the percentage of sentences containing the word in S-. On Hyland's list, words such as ought to, report, predict, propose, and assume are more likely to appear in S+ than in S-. On the scientific uncertainty list, words such as inconclusive, inconsistent, and uncertain have the highest S+/S- ratios. In particular, the word inconclusive with the highest S+/S- ratio clearly indicates a significant degree of uncertainty. Furthermore, its strong S+/S- ratio also suggests that contradictions are a major source of uncertainty and that using common hedge words alone may not adequately identify the nature of the uncertainty one is dealing with. Given the theoretical implications of irreconcilable intellectual conflicts, it is important to emphasize that uncertainty detection should cover a diverse range of uncertainties, especially the ones that in theory may alter the course of the development of a scientific specialty.

**Table 1. Frequencies of hedging words (Hyland 1996) and scientific uncertainty cue words in the two sets of MEDLINE sentences, indicating contradictions (i.e. S+) are an important source of uncertainty.**

| Hyland (1996) | S+ | S+ (%) | S- | S- (%) | S+/S- | Scientific Uncertainty Cues | S+ | S+ (%) | S- | S- (%) | S+/S- |
|---|---|---|---|---|---|---|---|---|---|---|---|
| **ought to** | 73 | 0.205 | 5 | 0.014 | 14.582 | **inconclusive** | 169 | 0.475 | 4 | 0.011 | 42.197 |
| **report** | 5982 | 16.817 | 435 | 1.224 | 13.734 | **inconsistent** | 137 | 0.385 | 9 | 0.025 | 15.203 |
| **predict** | 521 | 1.465 | 80 | 0.225 | 6.504 | **uncertain** | 243 | 0.683 | 21 | 0.059 | 11.557 |
| **shall** | 8 | 0.022 | 2 | 0.006 | 3.995 | **often** | 1024 | 2.879 | 151 | 0.425 | 6.773 |
| **propose** | 371 | 1.043 | 100 | 0.281 | 3.705 | **speculate** | 43 | 0.121 | 9 | 0.025 | 4.772 |
| **might** | 326 | 0.916 | 128 | 0.360 | 2.544 | **conclusive** | 60 | 0.169 | 14 | 0.039 | 4.280 |
| **assume** | 131 | 0.368 | 63 | 0.177 | 2.077 | **hypothesis** | 591 | 1.661 | 141 | 0.397 | 4.186 |
| **may** | 2179 | 6.126 | 1140 | 3.209 | 1.909 | **surpris -** | 39 | 0.110 | 11 | 0.031 | 3.541 |
| **seem** | 399 | 1.122 | 231 | 0.650 | 1.725 | **unexpected** | 25 | 0.070 | 9 | 0.025 | 2.774 |
| **suggest** | 1458 | 4.099 | 854 | 2.404 | 1.705 | **not conclusive** | 12 | 0.034 | 5 | 0.014 | 2.397 |
| **cannot** | 105 | 0.295 | 115 | 0.324 | 0.912 | **apparent** | 533 | 1.498 | 256 | 0.721 | 2.079 |
| **will** | 249 | 0.700 | 280 | 0.788 | 0.888 | **likely** | 180 | 0.506 | 95 | 0.267 | 1.892 |
| **should** | 320 | 0.900 | 394 | 1.109 | 0.811 | **generally** | 138 | 0.388 | 77 | 0.217 | 1.790 |
| **could** | 413 | 1.161 | 511 | 1.438 | 0.807 | **not consistent** | 7 | 0.020 | 5 | 0.014 | 1.398 |
| **must** | 166 | 0.467 | 216 | 0.608 | 0.768 | **implying** | 5 | 0.014 | 5 | 0.014 | 0.999 |
| **indicat-** | 566 | 1.591 | 870 | 2.449 | 0.650 | **believe** | 67 | 0.188 | 71 | 0.200 | 0.942 |
| **would** | 148 | 0.416 | 229 | 0.645 | 0.645 | **implies** | 9 | 0.025 | 10 | 0.028 | 0.899 |
| **appear** | 410 | 1.153 | 971 | 2.733 | 0.422 | **suspect** | 35 | 0.098 | 40 | 0.113 | 0.874 |
| **could not** | 27 | 0.076 | 99 | 0.279 | 0.272 | **unlikely** | 11 | 0.031 | 16 | 0.045 | 0.687 |
| | | | | | | **probably** | 127 | 0.357 | 243 | 0.684 | 0.522 |
| | | | | | | **certain** | 130 | 0.365 | 311 | 0.875 | 0.417 |
| | | | | | | **presumably** | 16 | 0.045 | 60 | 0.169 | 0.266 |
| | | | | | | **cannot exclude** | 2 | 0.006 | 0 | 0.000 | n/a |

Figure 3 shows a log-transformed frequency-frequency diagram of the Hyland-1996 hedging words in S+ and S- sentences. Given a hedging word *w*, it is shown at (*log*($f_{S+}(w)$), *log*($f_{S-}(w)$)) in the diagram, where $f_{S+}(w)$ is the word frequency in S+ and $f_{S-}(w)$ is the word frequency in S-. The dashed red line marks where the occurrences of a word in S+ and S- are the same. Words located above the line have a strong presence in S-, whereas words below the line are typically found in S+.

Typical hedge words such as *may* and *suggest* occurred frequently in both S+ and S-. Words such as *report*, *often*, *hypothesis*, and *predict* occurred frequently in both S+ and S-, but relatively more in S+. The word *uncertain* is frequent in S+ and rare in S-. In contrast, the word *certain* is frequent in S- but moderate in S+. Words such as *inconclusive* and *inconsistent* occur relatively frequent in S+, but very rare in S-.

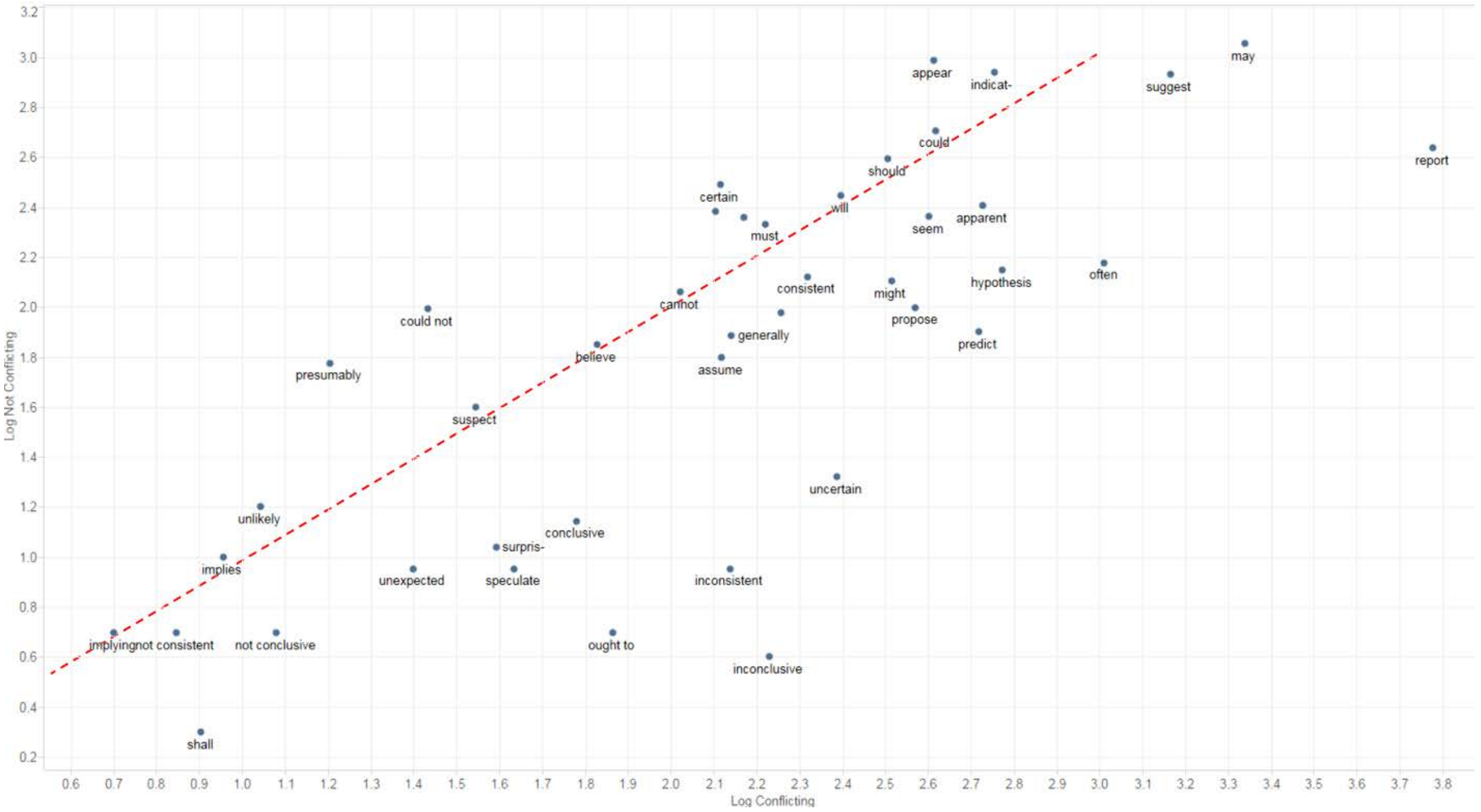

**Figure 3. A log-log frequency diagram of hedge words in S+ and S-. Hedge words are based on Hyland (1996).**

Table 2 presents some concrete examples of sentences in S+ to demonstrate the kinds of uncertainty due to conflicting information. For each sentence, the PubMed ID (PMID) of the source article is provided. The reader can retrieve the article by its PubMed ID (PMID) and explore the original context of the sentence. These examples illustrate patterns of uncertainty cue words such as "conflicting observations|results|information" or "evidence|finding is conflicting." The symbol "|" means "or."
Conflicting, contradictory, and surprising information may reflect a gap, mismatch, or bias between our current beliefs and the true value of a proposition under investigation. As shown in these examples, conflicting information provides an explicit explanation of the type of uncertainty involved. In contrast, if we focus on hedging alone such as "may suggest," we may miss the opportunity to learn more about the significance of the uncertainty.

**Table 2. Examples of sentences involving uncertainties due to conflicting findings.**

| PMID | Sentence |
|---|---|
| 6788027 | These conflicting observations may suggest the existence of two molecular species demonstrating NGF-like activity: one sharing antigenic determinants with mouse 2.5S NGF and the other antigenically unrelated. |
| 12221200 | These conflicting results may suggest that the cholesterol-lowering activity of products rich in oat beta-glucan depends on factors, such as its viscosity in the gastrointestinal tract, the food matrix and/or food processing. |
| 11360725 | This may suggest that N270 represented the response of the brain to conflicting information between different cortical levels. |
| 20501486 | The evidence for the effectiveness of compulsion in community mental health care is patchy and conflicting, with randomized or other trials failing to show significant benefits overall even if secondary analyses may suggest positive outcomes in some subgroups. |
| 23667851 | These clinical features may suggest a relatively weak DNE of A189Vcompared to other TP53 mutations, and in silico predictions and in vitro findings of the function of A189V mutant protein are conflicting. |

Table 3 includes four sentences and associated semantic predications from Semantic MEDLINE to illustrate the complexity and challenges of identifying uncertainties of propositions. In these examples, semantic predications serve the role of propositions because their truthfulness, or their uncertainty, is our interest. Each sentence in Table 4 has a sentence ID (SID) along with the PubMed ID (PMID) of the article it belongs to. Each predication has a PID number. These identifiers are included in Semantic MEDLINE. One can retrieve these records from Semantic MEDLINE through their corresponding identifiers.
The first sentence (SID 40120058) indicates the uncertainty of the prognosis of a surgical treatment of pancreatic cancer and the uncertainty is due to conflicting results. The semantic predication, i.e. the proposition (PID 686672), states "Excision TREATS Pancreatic carcinoma." The negation "has not been clearly defined" may provide a vague and weak indication of the uncertainty. In contrast, "conflicting results" is a much stronger and specific signal of the degree of uncertainty. Similarly, in the second sentence, "conflicting results" provides a clear indication of uncertainty regarding a surgical intervention's effect. The only other sign of uncertainty in the sentence is in the form of weasels (Vincze, 2013), i.e. the passive "have been reported" as opposed to a direct attribution of who has reported. However, the phrase "have been reported" is routinely used in scientific publications. If it is used as an uncertainty cue, it is likely to generate many false positives.
The complexity is even higher in the third and fourth examples. Both cases contrast something that is still unknown to something that is established. In both cases, "conflicting results have been reported" conveys the presence of uncertainty. In the third sentence, the epistemic status of the two propositions is established, but the focus of the sentence is on something different, i.e. a proposition that is specifically about AIDS patients. Similarly, in the fourth sentence, the status of propositions on mouse T cells is well established, the question is about propositions on human T cells. In both cases, it becomes necessary to introduce new propositions because existing propositions do not represent the research questions.

**Table 3. Examples of sentences and uncertainties of explicit and implicit propositions.**

| PMID | | | **Sentence** |
|---|---|---|---|
| 8116075 | SID | 40120058 | The prognosis after surgical resection for pancreatic cancer has ***not been clearly defined*** because *conflicting results* have been reported. |
| | PID | 686672 | Excision TREATS Pancreatic carcinoma |
| 6172161 | SID | 32504780 | *Conflicting results* have been reported on the ***influence*** of portacaval anastomosis on liver carcinogenesis. |
| | PID | 1720527 | Portacaval Shunt, Surgical AFFECTS Hepatocarcinogenesis |
| 8534426 | SID | 38710648 | In contrast to the established ***role*** of Helicobacter pylori gastritis in gastritis and duodenal ulcer in general, *conflicting results* have been reported in *patients with human immunodeficiency virus (HIV) infection and the acquired immunodeficiency syndrome*. |
| | PID | 4725669 | Helicobacter-associated gastritis AFFECTS Gastritis |
| | PID | 4725698 | Helicobacter-associated gastritis AFFECTS Duodenal Ulcer |
| 8765032 | SID | 53686452 | The ***role*** of interleukin-4 (IL-4) in the induction of IL-4 in mouse T cells is well established, but *conflicting results* have been reported with anti-CD3-primed *human T cells and T cell clones*. |
| | PID | 3363893 | Interleukin-4 AFFECTS T-Lymphocyte |
| | PID | 1081542 | T-Lymphocyte PART_OF House mice |

The above examples are only a small sample of possible scenarios in which information on the truthfulness of a proposition is either missing, insufficient, or questionable, hence the uncertainty. How can we expand this list systematically such that both commonly used hedge words and cue words of uncertainties deeply rooted in scientific inquiries are taken into account? We first compile a seed list to reflect the uncertainties identified in our conceptual framework, then expand the list computationally as detailed below.

### *Compiling a Seed List of 61 Cue Words*

The construction of the seed list aims to take into account uncertainties implied or expressed in a statement of a scientific proposition. We are particularly interested in uncertainties due to incomplete, inconsistent, or contradictory information as well as uncertainties hinted by hedging, speculation, or other indirect sources because inconsistencies in science may profoundly impact the epistemic status of a large number of propositions. Scientific uncertainties in controversies often have practical implications on decisions and policies on the public.

In addition to the hedging words suggested by Hyland (1996), we manually generated a set of **61** cue words of uncertainty based on a thesaurus of English. Then we searched frequencies of these words in several widely known resources of scientific publications, notably Google Scholar (excluding patents), ScienceDirect (journals only), the Web of Science (1980-2016/4/9), Springer (https://link.springer.com/), Mendeley, PubMed, core.ac.uk (English), US patents (full text since 1976 in USPTO), Supreme Court decisions (61,509 cases), and general-purpose documents (Google). The frequency of the word *knowledge* in each collection serves as a baseline in that the score of an uncertainty cue word is relative to the frequency of the word *knowledge*. For example, the word *unknown* has a score of 0.990 in Google Scholar, which means that the ratio of the frequency of the word *unknown* to the frequency of the word *knowledge* is 0.990. Since the word *knowledge* is very common in scientific publications, the relative score of a cue word provides a simple measure of its popularity.

As shown in Table 4, the word *unknown* is the most frequently used word from our list in all seven sources of scientific documents, except it ranks the 2nd in the Core.ac.uk collection after the word *uncertainty*. Other top-ranked cue words of uncertainty include *incomplete*, *conflicting*, *unusual*, and *unexpected*. Tackling the unknown is central to science. The word *uncertainty* is among top 10 on Google Scholar, ScienceDirect, Web of Science, Springer, and Core.

**Table 4. Top-10 most frequently used uncertainty cue words relative to the word *knowledge* in corresponding collections.**

| Google Scholar | | ScienceDirect | | Web of Science | | Springer | | Mendeley | | Pubmed | | Core | |
|---|---|---|---|---|---|---|---|---|---|---|---|---|---|
| unknown | 0.990 | unknown | 0.6021 | unknown | 1.3294 | unknown | 0.3416 | unknown | 0.3875 | unknown | 0.6526 | uncertainty | 0.4748 |
| incomplete | 0.755 | conflicting | 0.3613 | unusual | 0.5966 | conflicting | 0.2516 | unclear | 0.2166 | undetermined | 0.6203 | unknown | 0.4224 |
| impossible | 0.7251 | supris* | 0.3309 | suspect | 0.5217 | contrary | 0.1998 | uncertainty | 0.1806 | unclear | 0.3784 | impossible | 0.3895 |
| consensus | 0.6992 | contrary | 0.3308 | supris* | 0.4091 | impossible | 0.1972 | consensus | 0.1582 | unusual | 0.3269 | supris* | 0.3685 |
| uncertainty | 0.6952 | uncertainty | 0.3225 | uncertainty | 0.3527 | supris* | 0.1783 | unusual | 0.1376 | consensus | 0.2398 | contrary | 0.3226 |
| unexpected | 0.6394 | unclear | 0.2945 | controversial | 0.3373 | unclear | 0.1625 | contrary | 0.1187 | uncertain | 0.1965 | consensus | 0.2473 |
| supris* | 0.6016 | impossible | 0.2901 | contrary | 0.3341 | uncertainty | 0.1552 | controversial | 0.1121 | controversial | 0.1747 | incomplete | 0.2440 |
| uncertain | 0.5896 | suspect | 0.2426 | unclear | 0.2754 | incomplete | 0.1312 | incomplete | 0.1020 | incomplete | 0.1490 | ambigu{ity\|ous} | 0.2266 |
| unusual | 0.5319 | incomplete | 0.234 | conflicting | 0.2712 | suspect | 0.1296 | uncertain | 0.0979 | contrary | 0.1403 | unusual | 0.2164 |
| contrary | 0.5259 | unusual | 0.2285 | unexpected | 0.2401 | consensus | 0.1189 | unexpected | 0.0735 | conflicting | 0.1200 | inconsistent | 0.2039 |

In contrast to the uncertainty cue words' distributions in scientific texts, Table 5 shows their distributions in other resources, namely the US Supreme Court opinions, which include decisions reached after detailed arguments and justifications based on various evidence, USPTO, NSF awards' abstracts, and two general sources New York Times and the Google search engine. The overall distributions are different from collections of scientific publications. Supreme Court opinions featured words such as *contrary*, *controversial* and *dispute*. The USPTO and NYTimes highlight the word *impossible*. Interestingly, the word *uncertainty* is on the top of the list for NSF, which echoes our expectations regarding the central role of understanding *uncertainty* in science.

**Table 5. Occurrences of uncertainty cue words in resources rather than scientific publications.**

| Supreme | | USPTO | | NYTimes | | Google | | NSF | |
|---|---|---|---|---|---|---|---|---|---|
| contrary | 1.3001 | impossible | 1.1443 | impossible | 1.0749 | unknown | 0.9633 | uncertainty | 0.1980 |
| controversial | 1.0170 | contrary | 1.1167 | unusual | 0.8905 | supris* | 0.6358 | unusual | 0.1161 |
| dispute | 0.9794 | unknown | 0.7607 | dispute | 0.7309 | dispute | 0.6064 | debatable | 0.0868 |
| inconsistent | 0.7455 | incomplete | 0.3731 | contrary | 0.5375 | myster* | 0.5284 | conflicting | 0.0817 |
| impossible | 0.4704 | unexpected | 0.3715 | unknown | 0.5128 | impossible | 0.3972 | surpris* | 0.0672 |
| ambigu* | 0.2825 | supris* | 0.3029 | suspect | 0.3472 | unusual | 0.2358 | incomplete | 0.0540 |
| conflicting | 0.2678 | unusual | 0.2380 | unexpected | 0.3306 | unexpected | 0.1771 | uncertain | 0.0516 |
| doubtful | 0.2226 | incompatible | 0.1970 | uncertain | 0.3192 | suspect | 0.1752 | impossible | 0.0510 |
| unusual | 0.2178 | inconsistent | 0.1898 | suspicion | 0.3004 | bizarre | 0.1523 | unexpected | 0.0400 |
| uncertainty | 0.1844 | unreliable | 0.1545 | controversial | 0.2822 | controversial | 0.1431 | consensus | 0.0388 |

We used Principal Component Analysis (PCA) to group uncertainty cue words in the seed list to seven dimensions based on their distribution across the 12 text collections (See Table 6). These dimensions represent the types of uncertainties in collections of scientific documents. The primary component contains words such as *misleading*, *fallacy*, *incomprehensive*, *uncertain*, and contradictory. The second group contains *dispute*, *doubtful*, *unconvincing*, and *controversial*. The third one contains *mysteries*, *bizarre*, and *skeptical*.

**Table 6. Top 10 words on each of the seven PCA components derived from the word by source matrix.**

| F1 | | F2 | | F3 | | F4 | | F5 | | F6 | | F7 | |
|---|---|---|---|---|---|---|---|---|---|---|---|---|---|
| misleading | 0.96 | dispute | 0.84 | myster{ious\|y\|ies} | 0.77 | unreliable | 0.54 | surpris{ing\|e} | 0.44 | unknown | 0.61 | undetermined | 0.81 |
| fallacy | 0.93 | doubtful | 0.75 | bizarre | 0.73 | incompatible | 0.51 | skeptic | 0.41 | suspect | 0.59 | unrecognized | 0.31 |
| incomprehensible | 0.93 | unconvincing | 0.60 | uncharted | 0.66 | impossible | 0.49 | uncharted | 0.37 | contrary | 0.44 | unclear | 0.31 |
| uncertain | 0.93 | irreconcilable | 0.59 | undiscovered | 0.52 | contrary | 0.42 | misconception | 0.36 | controversial | 0.37 | | |
| perplexity | 0.92 | inconceivable | 0.56 | unknown | 0.48 | unanticipated | 0.42 | parado{xical\|x} | 0.35 | unexplained | 0.31 | | |
| contradictory | 0.91 | controversial | 0.53 | baffling | 0.39 | surpris{ing\|e} | 0.40 | misbelief | 0.32 | conflicting | 0.31 | | |
| flaw | 0.91 | deceptive | 0.49 | surpris{ing\|e} | 0.39 | unpredictable | 0.38 | ambigu{ity\|ous} | 0.31 | | | | |
| contentious | 0.89 | suspicion | 0.49 | skeptic | 0.37 | uncharted | 0.34 | controversial | 0.31 | | | | |
| incongruity | 0.89 | improbable | 0.46 | unusual | 0.36 | myster{ious\|y\|ies} | 0.32 | implausible | 0.19 | | | | |
| unexpected | 0.87 | skeptic | 0.45 | suspect | 0.34 | undiscovered | 0.32 | irreconcilable | 0.03 | | | | |

We mapped the 12 collections of text documents based on distributions of uncertainty cue words (Figure 4). Supreme Court decisions is an outlier. The most similar ones are Mendeley, Springer, Core, and ScienceDirect, which are all full-text collections. Pubmed and NSF award abstracts are their nearest neighbors. Google, the Web of Science (WoS), and Google Scholar appear in the lower half of the plot. New York Times and USPTO are relatively closer than others.

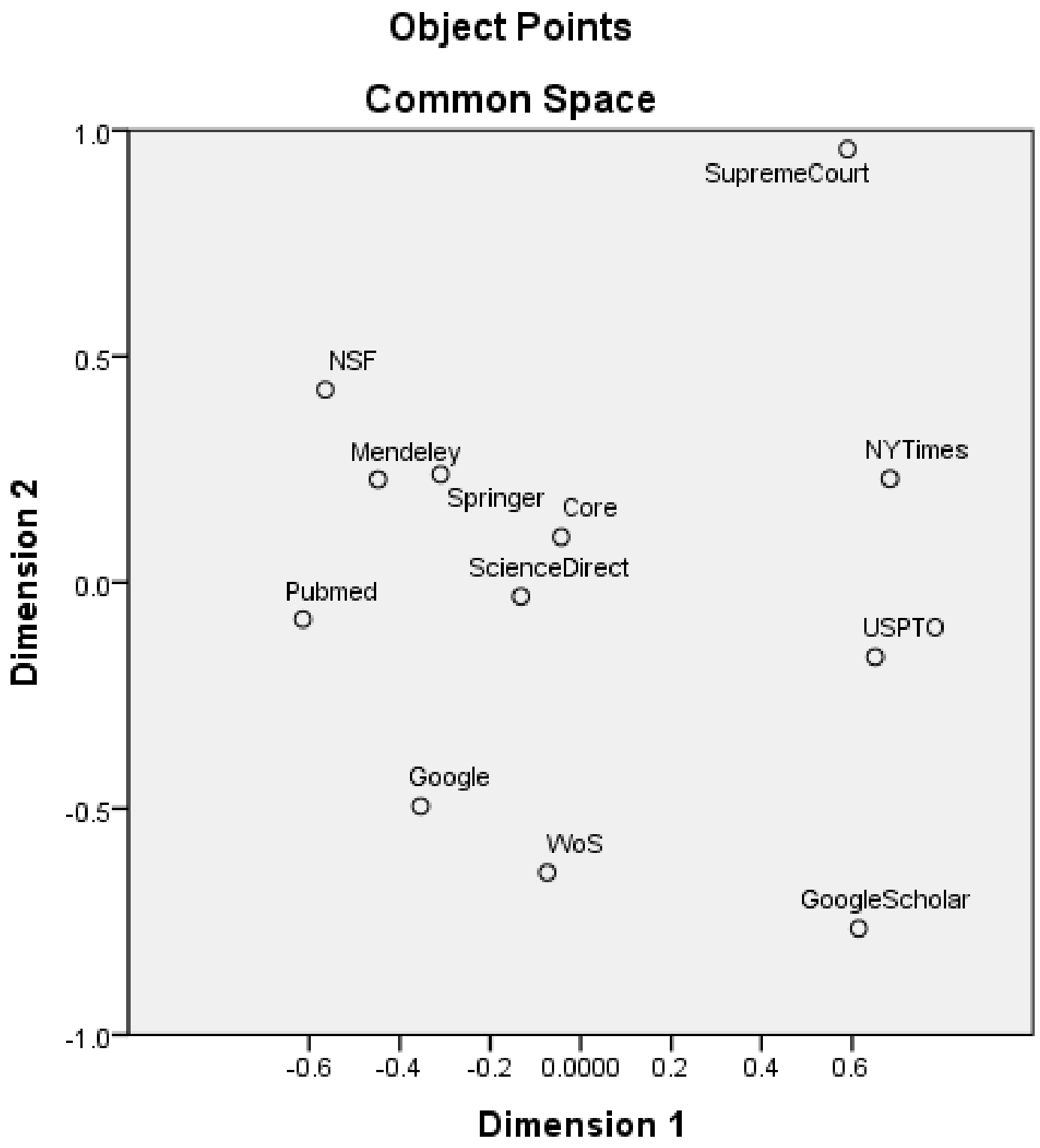

**Figure 4. A multidimensional scaling (MDS) configuration of the 12 text collections. Minkowski distance is used.**

### *Uncertainties of 24 Disciplines*

The use of hedging varies across scientific disciplines (Hyland, 1998; Hyland, 2006). For example, Dahl (2008) found hedging is more frequently used in linguistics than economics. Hu and Cao (2015) explained disciplinary influences of the use of hedging in terms of a theory that divides scientific disciplines into knowledge-dominated and knower-dominated ones (Maton, 2000). Disciplines dominated by a knowledge code, including many natural sciences, have established scientific principles and procedures to verify scientific publications and findings. In contrast, disciplines such as the humanities depend more on the distinct individual characteristics of those constructing disciplinary knowledge. Scientists, or “knowers”, carry more weights in these disciplines. The knowledge-knower distinction is influenced by the vertical and horizontal discourse structures proposed by Bernstein (1999).

We demonstrate how research on uncertainty can open up new ways to characterize the stability of a subject area. Using the Consyn database[1], Elsevier’s content syndication system, we estimate the overall uncertainty of a subject area in terms of the proportion of its publications containing uncertainty cue words. We expect that more epistemologically focused uncertainty cue words such as contradictions and conflicting results will provide more insights into a subject area than using hedging words alone.

Given a list of uncertainty cue words, we can see how often these words are used in a particular scientific discipline. For each discipline, we searched for articles in Consyn that contain at least one of the five words: *conflicting*, *contradictory*, *inconsistent*, *discrepant*, and *irreconcilable*. These words indicate situations where scientists cannot reach a consensus. The truthfulness of propositions involved in such situations is unsettled. Therefore, these words are indicators of underlying uncertainty, especially in the context of scientific inquiry because in non-scientific contexts one may settle with contradictions, whereas in science contradictions motivate further investigations rather than terminate a line of research. In such situations, the uncertainty is about what one can expect. For example, the uncertainty about what caused the mass extinctions 65 million years ago was high when there were over 80 competing theories. After the discovery of conclusive evidence for the impact theory, the overall uncertainty of the research topic reduced dramatically and researchers searched for new topics to study elsewhere (Chen, 2006; French & Keoberi, 2010).

[1] https://consyn.elsevier.com

We estimated the rate of uncertainty as the number of items found in each subject area on Consyn divided by the total number of items in the same area. For example, Psychology has a total of 220,250 items at the time of search, of which 70,096 items matched the five-uncertainty-cue query, thus the rate of uncertainty in Psychology is 32%. The rate represents a lower bound of the degree of uncertainty associated with the publications in a subject area. This is a rough estimate. Its accuracy may be improved by using an enriched list of uncertainty cues like the one we will introduce shortly. The rate of uncertainty words various across subject areas, which makes this an interesting topic to investigate in its own right.

All subject areas are divided into 5 groups based on their recalls (Table 7). The group with the highest rates of uncertainty includes psychology (32%), business, management, and accounting (28%), social sciences (26%), economics, econometrics, and finance (25%), and neuroscience (23%). The second group, with the rates between 18-20%, includes medicine and dentistry (20%), pharmacology, toxicology and pharmaceutical science (18%), and arts and humanities (18%). The third group, with the rates between 13-17%, includes environmental sciences (17%), immunology and microbiology (16%), and computer science (13%). The fourth group contains disciplines with rates between 8-12%, such as decision sciences (12%), engineering (9%), and energy (8). The fifth group, the lowest rates of all 4-7%, includes mathematics (7%), material science (4%), and chemistry (4%).

The above distribution of the uncertainty rates across disciplines may guide us further in prioritizing disciplines to study uncertainties. For example, with a rate of (7%), mathematics may not cover the entire spectrum of the scenarios of how uncertainties are present and how they are reconciled subsequently. In contrast, psychology with the rate of 32% is rich in terms of the variety of uncertainty types and instances. Depends on their needs, researchers may choose a subject area with a high or low rate of uncertainty cues.

**Table 7. The rate of uncertainty in a subject area.**

| 1 | Subject Area ( as of 8/13/2015 | Journal Items Only | Subtotal Items in Are | Rate % |
|---|---|---|---|---|
| 2 | Psychology | 70,096 | 220,250 | 32 |
| 3 | Business, management and accounting | 26,717 | 97,083 | 28 |
| 4 | Social sciences | 74,835 | 283,598 | 26 |
| 5 | Economics, econometrics and finance | 27,920 | 113,083 | 25 |
| 6 | Neuroscience | 99,908 | 434,270 | 23 |
| 7 | Medicine and Dentistry | 423,391 | 2,093,102 | 20 |
| 8 | Veterinary science and veterinary medicine | 24,390 | 126,768 | 19 |
| 9 | Pharmacology, toxicology and pharmaceutical science | 56,441 | 305,601 | 18 |
| 10 | Nursing and heal professionals | 39,692 | 218,124 | 18 |
| 11 | Arts and humanities | 14,470 | 78,844 | 18 |
| 12 | Environmental Sciences | 56,594 | 328,192 | 17 |
| 13 | Immunology and microbiology | 51,184 | 310,404 | 16 |
| 14 | Agricultural and biological sciences | 63,010 | 400,272 | 16 |
| 15 | Biochemistry, genetics and molecular biology | 120,012 | 800,766 | 15 |
| 16 | Computer science | 32,040 | 252,366 | 13 |
| 17 | Decision sciences | 17,500 | 144,119 | 12 |
| 18 | Earth and Planetary Sciences | 24,393 | 225,816 | 11 |
| 19 | Engineering | 45,281 | 510,624 | 9 |
| 20 | Energy | 18,253 | 235,489 | 8 |
| 21 | Mathematics | 17,737 | 239,676 | 7 |
| 22 | Physics and astronomy | 28,507 | 498,418 | 6 |
| 23 | Chemical engineering | 17,434 | 355,512 | 5 |
| 24 | Material science | 24,038 | 608,991 | 4 |
| 25 | Chemistry | 20,585 | 522,442 | 4 |

## *Expansion and Prediction of Uncertainty Cue Words*

In this part of the study, we expand the seed list of 61 uncertainty cue words with a machine learning approach and validate the expanded list by two evaluators. Then we visualize the expanded words in the context of the seed list to demonstrate what we have gained from the expansion process.

Word2Vec is a group of two-layer neural network models for word embeddings (Mikolov et al., 2013). Word2Vec takes a large corpus of text as input and produces a vector space of several hundred dimensions. Each word in the corpus is represented by a vector in the space. Words that share common contexts in the corpus are located in close proximity to one another. In other words, if words are used in similar contexts, they are similar in the vector space. Thus, among other applications, Word2Vec can be used to find semantically equivalent words to words on our seed list.

We considered two Word2Vec models. One is the Google News Word2Vec model[2] and the other is the PubMed Word2Vec model. In a Word2Vec model, each word consists of a context vector based on either skip-gram with the PubMed Word2Vec or Continuous Bag of Word (CBOW) with the Google News Word2Vec model. Given a word, these models can identify words that tend to be used in similar contexts. For example, using uncertainty cue words such as 'inconsistent' as a query and limiting the output to the top 50 most similar words would identify 2,820 and 2,826 pairs of words from the PubMed and the Google News word2vec model, respectively. Table 8 shows top 10 words that are closely related to the word 'inconsistent' in the PubMed model. These words are considered as candidates for expansion.

**Table 8. Semantically equivalent words to "inconsistent." Words are case-insensitive.**

| Candidate words | similarity |
|---|---|
| contradicting | 0.71122 |
| consistent | 0.66428 |

[2] https://code.google.com/archive/p/word2vec/

| | |
|---|---|
| Inconsistent | 0.65610 |
| disappointing | 0.63907 |
| equivocal | 0.62048 |
| discrepant | 0.61734 |
| Contradictory | 0.59753 |
| encouraging | 0.59554 |
| contradicted | 0.59220 |
| Conflicting | 0.58962 |

This unique feature of Word2Vec models enables the identification of semantically equivalent words to a given uncertainty cue word. In particular, we utilized two Word2Vec models to search for candidate cue words, namely the Google News model and the PubMed model. The PubMed Word2Vec model was built based on 23 million PubMed records with the skip-gram learning algorithm to create 200-dimensional vectors using a window size of 5, hierarchical softmax training, and a frequent word subsampling threshold of 0.001 (Pyysalo et al., 2013). The PubMed model consists of 2.35 million words. In contrast, the Google Word2Vec model was pre-trained with the CBOW algorithm on part of the Google News dataset containing 100 billion words (Mikolov et al., 2013). The model contains 300-dimensional vectors of 3 million words and phrases with sub-sampling using threshold of 1e-5 and negative sampling.

Using the two Word2Vec models, the seed list of uncertainty cue words is expanded as follows:

1) From each of the word2vec model, retrieve 50 most relevant terms to our initial cue words (61 words) and their semantic similarity scores.

The Google News model produced 2,826 pairs of similar words. The PubMed model produced 2,820 pairs . The number of distinct words retrieved from the Google News model is 2,151 and the number of distinct words from the PubMed model is 1,877.

2) For each of the seed cue words, computed the Pointwise Mutual Information (PMI) score for the retrieved term and the cue word. In addition, we computed the TF*IDF score for each word retrieved. Given the two word2vec models, we computed two sets of scores. PMI is a correlation of two events, x and y; The pointwise aspect of PMI indicates that we are considering specific events by the following formula: $pmi(x;y) = log(p(y|x)/p(y))$. TF*IDF is a well-received term weighting algorithm, introduced by Salton and his colleagues (1983) used in Information Retrieval and Text Mining.

3) Retain common words retrieved from both models. This step resulted in 393 distinct words as the candidate words for the expanded uncertainty cue words (See Supplementary Files).

4) The 393 expanded candidate words are reviewed by two evaluators independently. They rated whether a candidate word is valid cue word of uncertainty (Table 9).

**Table 9. Classifications of uncertainty words by two evaluators.**

| | Judge 1 | | |
|---|---|---|---|
| Judge 2 | Positive | Negative | Total |
| Positive | **151** | 49 | 200 |
| Negative | 63 | **130** | 193 |
| Total | 214 | 179 | 393 |

The two judges agreed on 151 words as valid cue words of uncertainty (positive) and agreed on 130 words that should be rejected (negative). The percentage of agreement is 281/(151+130) = 71.5%. Cohen’s kappa is 42.91% after taking into account the number of agreements that may occur purely by chance (Cohen, 1960). According to Landis and Koch’s (1977) interpretation, this value is ‘moderate’ in terms of strength of agreement.

We then chose the 151 expanded terms that both evaluators agreed on as the additional uncertainty cue words. We also include 130 negative cue words that both evaluators agreed on, which are not uncertainty words to build the training dataset for uncertainty cue word classification. The classification is binary since there are only two possibilities for each given word: either valid as a cue word or not. In order to include enough negative samples, we collected 100 unrelated terms to the uncertainty cue words from the Google News and PubMed models and combined these unrelated terms with correct ones to build a training dataset.

5) The classification of uncertainty cue words is evaluated with several machine learning algorithms, namely Recurrent Neural Network (RNN), k-Nearest Neighbors (KNN), Naïve Bayes, Random Forest, and Sequential Minimal Optimization (SMO).

- *Recurrent Neural Network (RNN)*: This is an increasingly popular deep learning algorithm. The key feature of an RNN is that the network contains at least one feedback connection, thus the activation flow forms a loop, which enables the network to perform temporal processing and learn sequences, e.g., perform sequence recognition/reproduction or temporal association/prediction (Schmidhuber, 2015).
- *k-Nearest Neighbors (KNN)*: KNN is a simple non-parametric machine learning algorithm. KNN does not make use of the training data points for generalization, which means there is no explicit or minimum training phase. This makes the training phase fast and makes decision based on the entire training data set (Altman, 1992)
- *Naïve Bayes*: Naive Bayes is a simple probabilistic machine learning algorithm by Bayes' theorem with the independence assumption among features. The independence assumption means that the value of a feature is independent of the value of any other features when the class variable is given (Hand & Yu, 2001).
- *Random Forest*: Random Forest is an ensemble machine learning algorithm that builds a multitude of decision trees at training time and generates the class that is the mode of the classes of the individual trees (Ho, 1995).
- *Sequential Minimal Optimization* (SMO): SMO is a variation of Support Vector Machine (SVM) algorithm to solve the training problem of SVM (Platt, 1998). SMO uses heuristics to partition the training problem into smaller problems that can be solved analytically.

In particular, RNN uses the following parameter setting: the number of channels is 6, the batch size is 100, the number of epochs is 500, and the number of iterations is 100. We also set the learning rate to be 0.0005 and chose stochastic gradient descent as the optimization algorithm for RNN. For the other four algorithms, we used the default setting provided in WEKA, a well-accepted machine learning tool (Witten et al., 2011). We used 10-fold cross-validation for the evaluation step, which is a technique for validating classification models by assessing how the outcome of a classification algorithm can generalize to an independent dataset. We also used standard performance measures such as accuracy, precision, recall, and F-measure.

Overall RNN outperformed the other four machine learning algorithms in terms of accuracy, precision and F-1, except for recall. SMO was the second best. KNN performed the worst (See Figure 5). Although RNN requires a long training time due to its characteristics of recurrent learning, the accuracy of its prediction is outstanding.

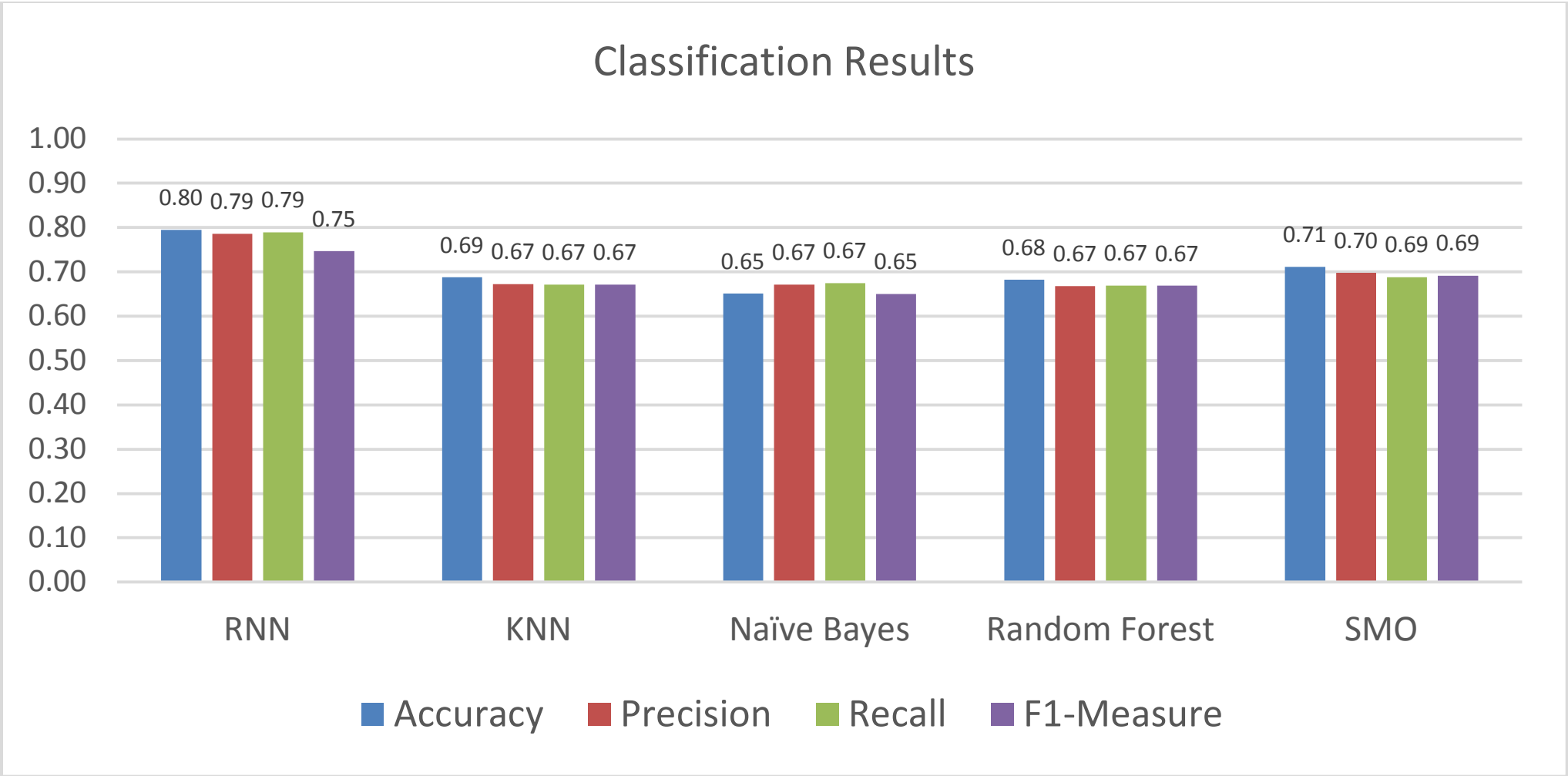

**Figure 5. Performance of machine-learning classifications of uncertainty cues.**

### *Visualization*

The Word2Vec-based expansion generated two networks of interrelated words. Connections between words are determined by their proximity in Word2Vec models. The Google News Word2Vec model generated a network of 435 interrelated words, whereas the PubMed Word2Vec model generated a network of 430 words. Words that are closely connected in these networks are semantically equivalent because they tend to appear in similar contexts.

In order to aggregate semantically equivalent cue words of uncertainty, we divided each network into clusters using the community detection algorithm by Blondel et al. (2008). Words in each cluster are more similar to one another than words in different clusters. The important of a node in a network can be measured with many metrics such as PageRank (Brin & Page, 1998), eigenvector centrality, and degree centrality. Since PageRank is particularly suitable for identifying cue words that are connected to other important cue words, we visualize the two networks, one from the Google News Word2Vec model and the other from the PubMed Word2Vec model, with Gephi, a network visualization tool and highlight important cue words based on their PageRank scores.

Figure 6 shows a visualized network of expanded uncertainty cue words based on the Google News Word2Vec model. The network contains three types of words, namely the 61 original seed words, 195 expanded words accepted by two judges, and candidate words rejected by the two judges. The label of a word *w* in the visualization is shown with the format *w* – *a* – *b* to reflect whether *w* is a seed word (*a* = 1 for yes, or 0 for no) and whether *w* is an accepted candidate word (*b* = 1 for yes, or 0 for no). For example, the label *paradox* – 1 – 1 means that the word *paradox* is a seed word, which is by definition accepted by judges as a valid cue word. In comparison, the label *inaccurate* – 0 – 1 means that the word *inaccurate* is not a seed word; instead, it is a new word from the Word2Vec model and it is an accepted by the judges as a valid uncertainty cue word. In contrast, the label *erroneous* – 0 – 0 means that the word *erroneous* is suggested by the model but rejected by the judges.

Words in the network are divided into 12 clusters based on the strengths of their connectivity. The largest four clusters are colored in red, green, blue, and purple, respectively. The size of the label of a word is proportional to its PageRank score, which means words with larger-sized labels are more important. The largest cluster, located on the right, contains prominent seed words such as *paradox* and *ambiguity*, newly expanded and accepted words such as *contradictions* and *indeterminacy*, and rejected candidates such as *dichotomy* and *duality*. It appears this cluster contains nouns mostly. These examples illustrate the effect of the expansion. Newly added cue words such as *contradictions* are semantically equivalent to seed words in the same cluster such as *paradox*, *fallacy*, and *ambiguity*. In addition, the visualization reveals that the word *contradictions* is closer to the seed word *ambiguity*, reflecting an influence of the Google

News as the input source for the Word2Vec model. Genre-dependencies of linguistic patterns have been addressed in the literature, e.g., by Szarvas et al. (2012).

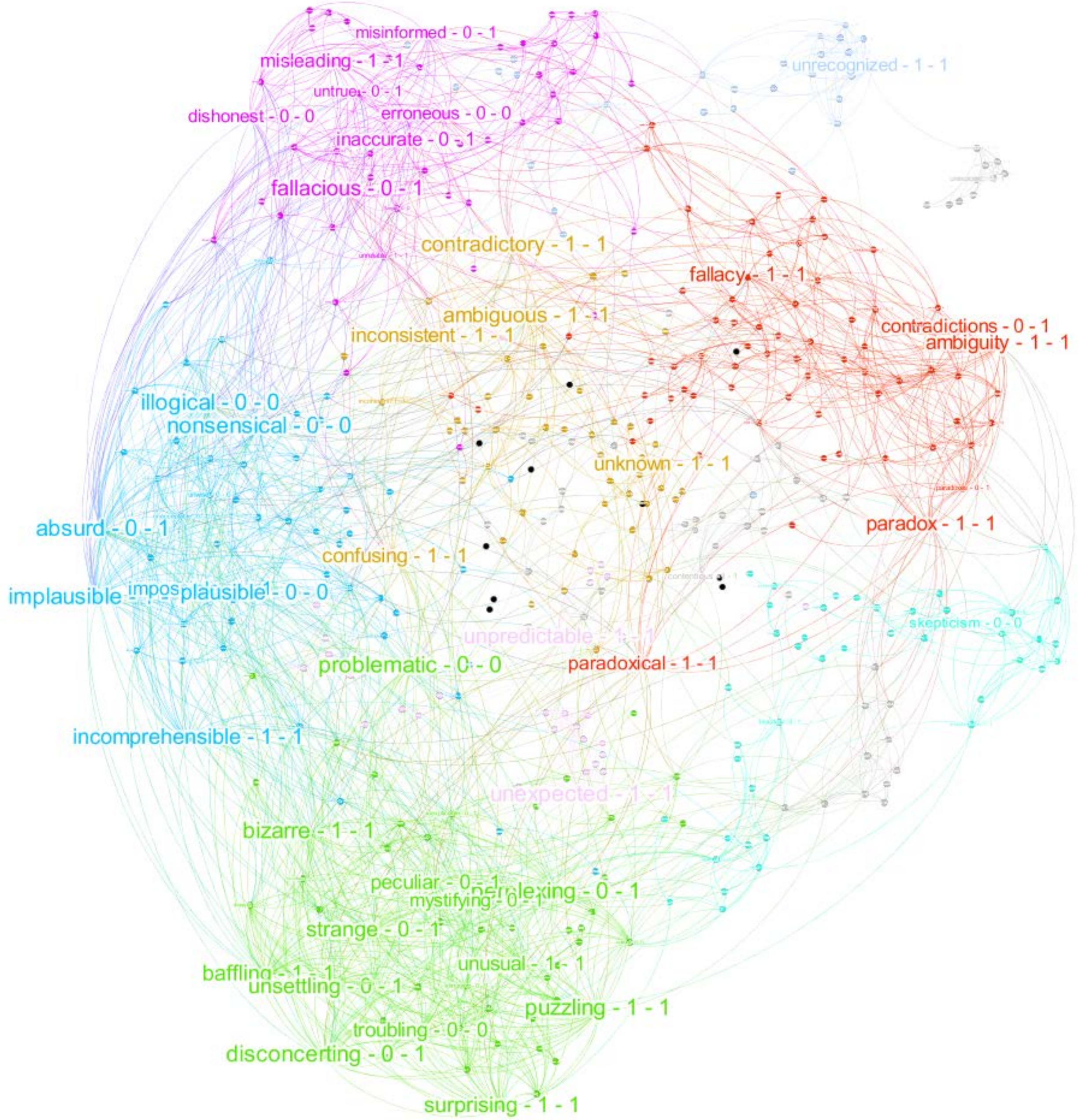

**Figure 6. A network of 425 words and 2,119 similarity links, including 61 seed words (* - 1 - *) and 193 confirmed expanded words (* - * - 1) based on the Google News Word2Vec model. Words are colored by 12 modularity groups. The label size of a word is proportional to its PageRank score.**

Table 10 includes illustrative examples of words from the largest four clusters from the Google News model (Clusters 6, 0, 5, and 3) and the PubMed model (Clusters 0, 5, 4, and 14). The values in the expanded column indicate whether the corresponding words are accepted by the judges as valid cue words of uncertainty. Because of the genre differences, the PubMed model is preferable for studying uncertainties in scientific domains, especially in biomedical domains, whereas the Google News model is preferable for studying uncertainties in mass media.

**Table 1. Examples of words in major clusters based on the Google News Word2Vec model.**

| | Google News | | | PubMed | | | |
|---|---|---|---|---|---|---|---|
| Word | Cluster | Seed | Expanded | Word | Cluster | Seed | Expanded |
| Paradox | 6 | Yes | Yes | Irreconcilable | 5 | Yes | Yes |
| Contradictions | 6 | No | Yes | Unsettling | 5 | No | Yes |
| Dichotomy | 6 | No | No | Dishonest | 5 | No | No |
| Surprising | 0 | Yes | Yes | Perplexity | 4 | Yes | Yes |
| Strange | 0 | No | Yes | Misunderstanding | 4 | No | Yes |
| Troubling | 0 | No | No | Ignorance | 4 | No | No |
| Implausible | 5 | Yes | Yes | Puzzling | 14 | Yes | Yes |
| Absurd | 5 | No | Yes | Uncommon | 14 | No | Yes |
| Illogical | 5 | No | No | Troubling | 14 | No | No |
| Uncertainty | 3 | Yes | Yes | Controversial | 0 | Yes | Yes |
| Skeptical | 3 | No | Yes | Questionable | 0 | No | Yes |
| Skepticism | 3 | No | No | Enigma | 0 | No | No |

Figure 7 shows the distributions of word types across clusters in the network. The distribution of the seed words, newly accepted words, and rejected words in each cluster provides several types of useful information. Which clusters do represent our own expertise in terms of the number of seed words? Where is the Word2Vec model's expertise in terms of the number of accepted words? For example, Clusters 2 and 6 contain most of our seed words, 12 and 10, respectively, suggesting that we may be particularly interested in these areas. Cluster 2 contains words such as *ambiguous* and *contradictions*, whereas Cluster 6 contains words such as *paradox*, *ambiguity*, *fallacy*, and *inconsistency*. In both clusters, the expansion generated as twice as many new cue words of uncertainty. In contrast, Cluster 7 contains five seed words, a relatively small number, but the expansion added 22 new cue words. The seed words in this cluster include *misleading*, *unreliable*, and *contrary*, whereas leading new words include *fallacious*, *inaccurate*, *misinformed*, *deceitful*, and *contradicting*. These examples suggest that hand-picked seed words may be biased due to individuals' preferences and prior experiences and the expansion method may compensate a potentially biased seed list by adding more semantically equivalent words that were not initially covered.

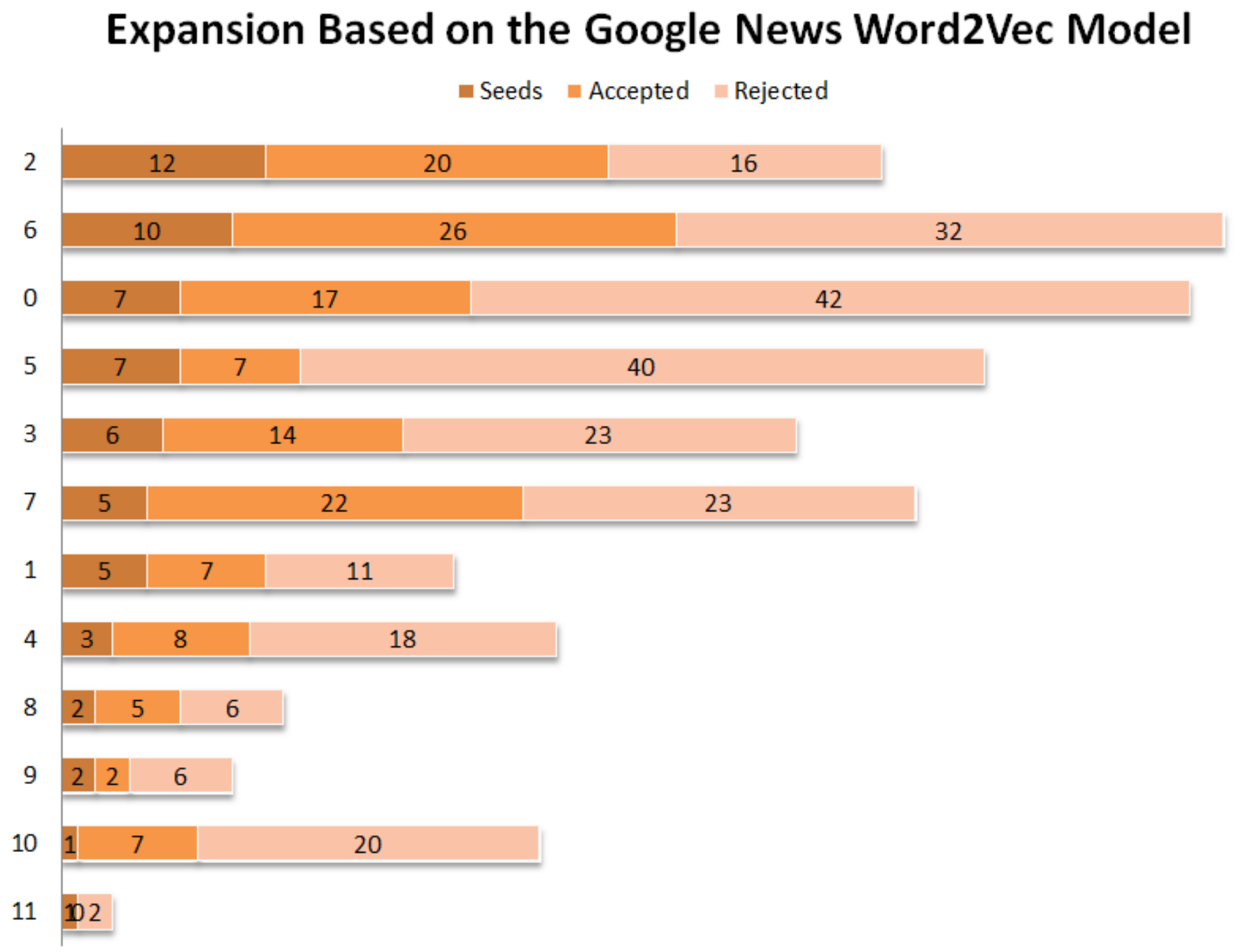

**Figure 7. Distributions of word types by cluster based on the expansion using the Google News model.**

Table 11 shows MEDLINE sentences retrieved based on uncertainty cue words. For each cluster, two sentences are chosen: one contains a seed cue word of uncertainty and the other contains an expanded cue word. The word *unproven* is an expanded cue word of uncertainty in Cluster 0. Similarly, words such as *unsettled* in Cluster 1, *absurdity* in Cluster 2, and *misguided* in Cluster 3 illustrate their validity in identifying sentences with uncertainty. For example, "*current medical documentation of dog bits may be misguided*" challenges the current status of the documented knowledge, which can be seen, in turn, as an indication of uncertainty of a previously accepted proposition. Note that the validation of a word was done independent of any concrete sentences. Sentences in the table may contain false positives.

**Table 2. Sample MEDLINE sentences containing uncertainty cue words.**

| Cluster | Cue Word | Seed | Accepted | PMID | Sentence |
|---|---|---|---|---|---|
| 0 | Unknown | Yes | | 18635088 | The source of the virus is **unknown** since it has not been detected in thin sections of intact hydra or in algal cells immediately after their isolation. |
| 0 | Unproven | No | Yes | 22432670 | We present a suspected but **unproven** case of MVEV infection to illustrate some of the challenges in clinical management. |
| 1 | Doubtful | Yes | | 1715963 | On the other hand, the relation of hepatitis C virus with sporadic acute non-A, non-B hepatitis may be **doubtful**. |
| 1 | Unsettled | No | Yes | 8972691 | While HTLV-I has been clearly associated with disease, the health implications of HTLV-II infection are still **unsettled**. |
| 2 | Paradoxical | Yes | | 24194956 | Chronic lymphocytic leukemia (CLL) is characterized by progressive hypogammaglobulinemia predisposing affected patients to a variety of infectious diseases but **paradoxically** not to cytomegalovirus (CMV) disease. |
| 2 | Absurdity | No | Yes | 27912859 | Additional sources of interest are the phenomenology of responsibility by Emmanuel Lévinas and works on **absurdity** and rebellion by Albert Camus. |
| 3 | Implausible | Yes | | | The current study investigated age-related differences in associative memory under conditions that were expected to differentially promote unitization, in this case by manipulating the spatial arrangement of two semantically unrelated objects positioned relative to each other in either spatially **implausible** or plausible orientations. |
| 3 | Misguided | No | Yes | 28398940 | Although accurate medical documentation of dog bites is a prerequisite to develop effective prevention strategies, current medical documentation of dog bites may be **misguided**. |
| 4 | Unpredictable | Yes | | 28302445 | Hence, investigation of the beneficial effects of agmatine on chronic **unpredictable** mild stress (CUMS) - induced depression, anxiety and cognitive performance with the involvement of nitrergic pathway was undertaken. |
| 4 | Tricky | No | Yes | 28044978 | Meanwhile, the analysis of the decision process induced by a nudge shows that it does not simply amount to a change in the environment and that its handling is ethically **tricky**. |
| 5 | Inconsistent | Yes | | 28402017 | Staff knowledge was higher in groups that had received asthma education, although results |
| | | | | | were **inconsistent** and difficult to interpret owing to differences between scales (low quality). |
| 5 | Misconstrues | No | Yes | 25080560 | There are two main problems with this approach: (1) constructing the debate over minimal risk as a disagreement between a uniform and a relative interpretation **misconstrues** the main difference between competing interpretations and (2) neither the uniform nor the relative interpretation identifies one unique and consistent group of children as the referent for minimal risk. |
| 6 | Bizarre | Yes | | 3184352 | **Bizarre** manifestations of VZV infection could present both diagnostic and therapeutic dilemmas. |
| 6 | Perplexing | No | Yes | 6248840 | Arthritis associated with coxsackievirus or adenovirus infection may be particularly **perplexing**, as the dominant syndrome may be a classic Still's variety of juvenile rheumatoid arthritis. |
| 7 | Consensus | Yes | | 20886705 | There was a mixture of **consensus** and mutant virus variants in the trachea and a mixture of mutant ones in the lung. |
| 8 | Unexpecting | No | No | 22939534 | **Unexpecting** age. |
| 9 | Debatable | Yes | | 9855375 | The role of HIV in PH is still **debatable**. |
| 9 | Disputing | No | Yes | 28334426 | To avoid making **disputing** assumptions on recurrent events or biomarkers after the failure event (such as death), the model is constructed on the basis of survivors' population. |
| 10 | Undetermines | No | No | 24241494 *No exact match. | These data demonstrate that expression patterns of circulating microRNAs are altered in multiple myeloma and monoclonal gammopathy of **undetermined** significance and miR-744 with let-7e are associated with survival of myeloma patients. |

In the PubMed-based expansion (Figure 8), there are 16 clusters. The largest four clusters, 0, 5, 4, and 14. Cluster 0, colored in purple and located at the bottom of the visualization, contains seed words such as *contentious*, *controversial*, and *uncertain* and accepted new words such as *unsettled*, *questionable*, *unexplored*, and *unresolved*. Cluster 5, colored in red and located at the top of the visualization contains seed words such as *incomprehensible* and *irreconcilable* as well as accepted words such as *muddled* and *unsettling*. Cluster 4, colored in green and located along the right-hand side of the visualization, featured seed words such as *perplexity*, *mysteries*, and *uncertainty* with accepted words such as *misunderstanding* and *inconsistency*. Cluster 14, colored in blue and located at the center of the graph, contains seed words such as *unrecognized*, *puzzling*, *confusing*, *unusual*, and *surprising* along with accepted cue words such as *troubling*, *misunderstood*, *uncommon*, and *unusual*.

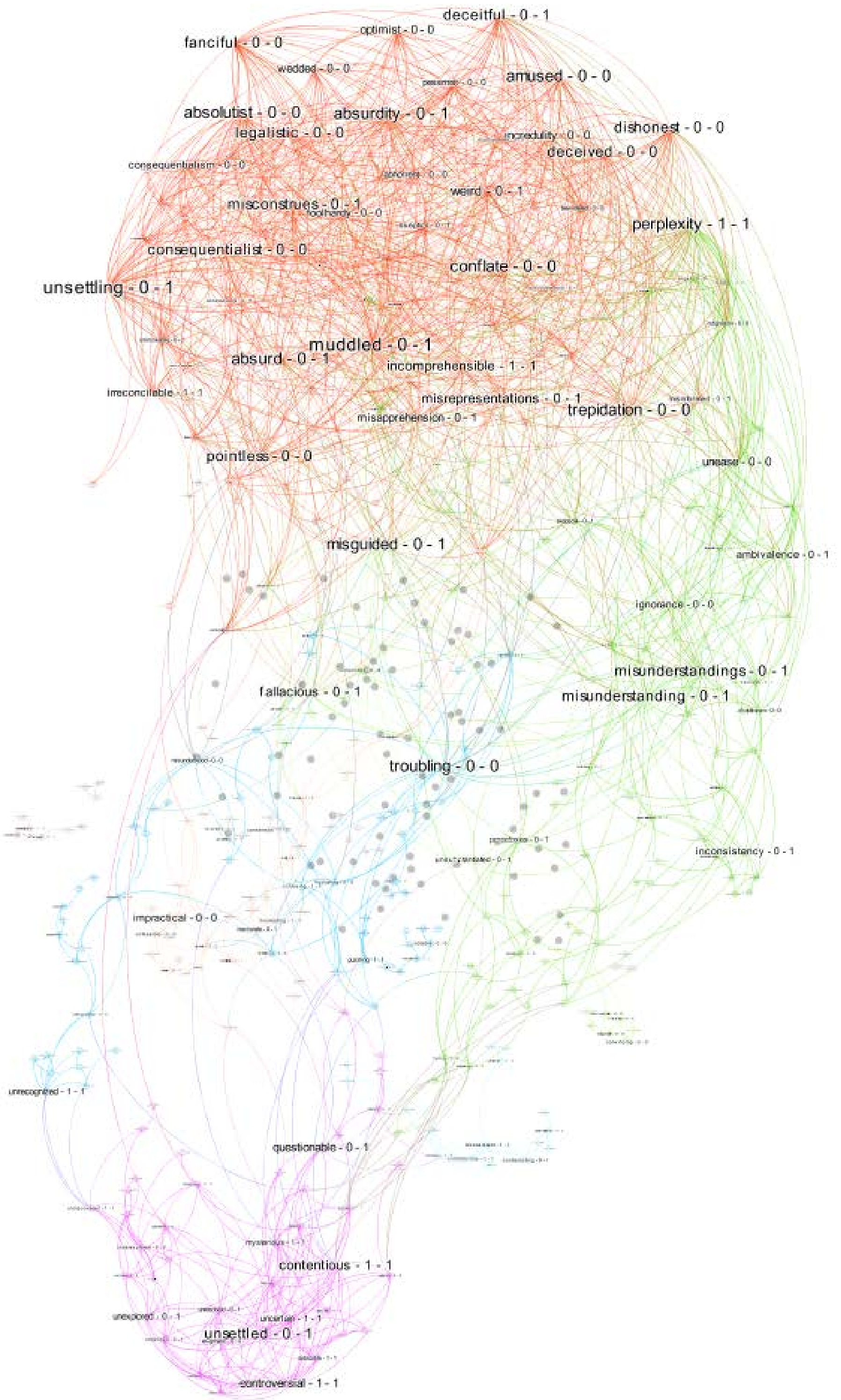

**Figure 8. A network of 435 words based on the PubMed word2vec model in 16 similarity clusters, including 61 seed words, 193 accepted candidates, and the remaining words are rejected candidates.**

The chart in Figure 9 shows the distributions of word types, i.e. seed, accepted, or rejected, based on the expansion using the PubMed Word2Vec model. Unlike with the Google News model, the majority of our seed words fall into Cluster 0, containing 24 seed words. The next group of seed words is in Cluster 4, containing 9 seed words. Clusters 5 and 14 contain five seed words each and gained new cue words by 4 and 3 times, respectively. The rest of the clusters are rather small.
Cluster 0 contains cue words such as *contentious*, *controversial*, and *uncertain*. The growth of this cluster from the expansion is significant, suggesting at least in the biomedical domain uncertainties due to the lack of information – *unsettled*, *unexplored*, *unresolved*, *undiscovered*, *unknown*, and *unaddressed* – play a central role in scientific discourses.

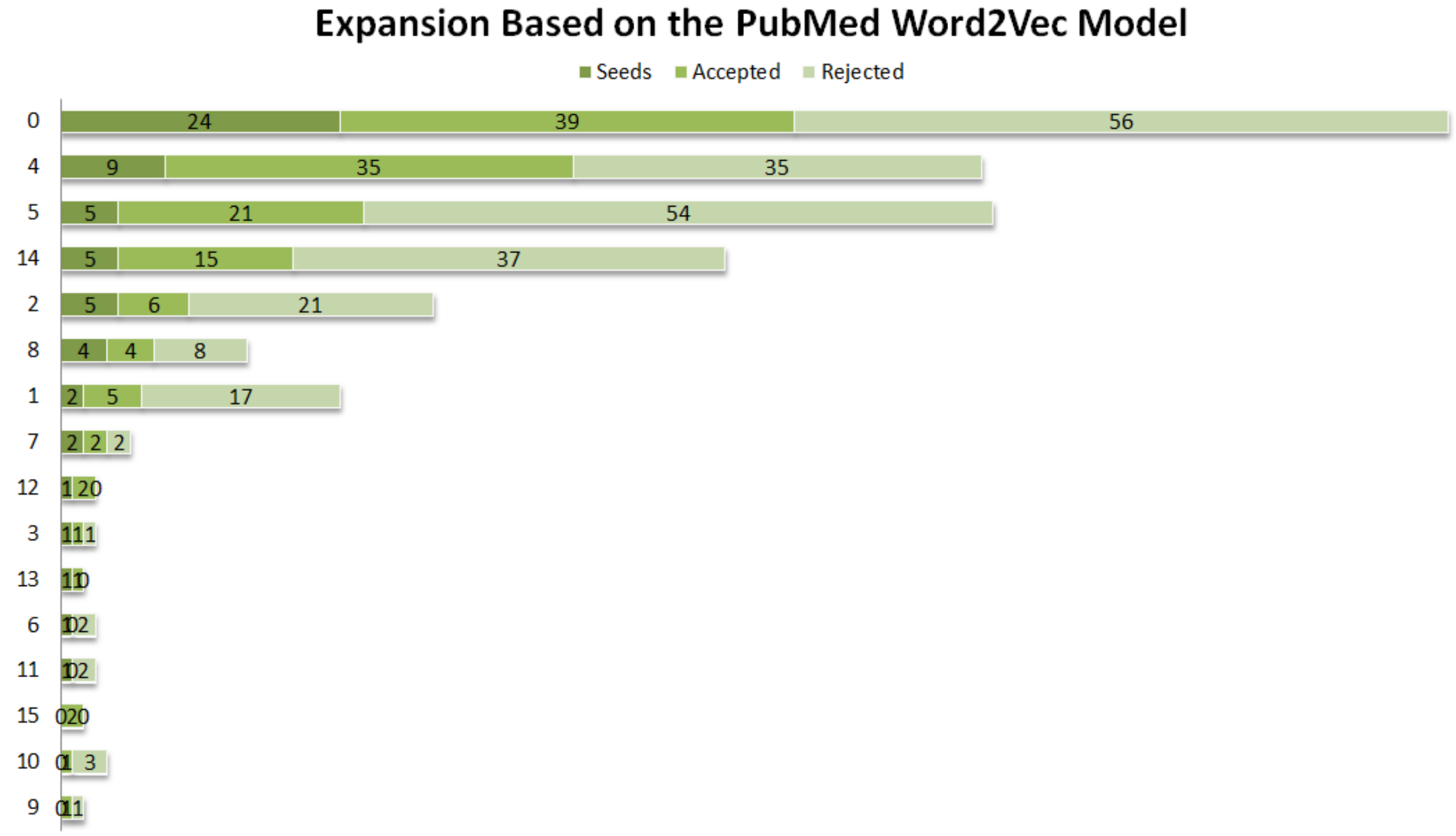

**Figure 9. Distributions of word types by cluster based on the expansion using the PubMed Word2Vec model.**

## Discussions and Conclusions

In this study, we introduced a conceptual framework for the study of uncertainties in scientific literature. The framework incorporates uncertainties hinted by hedge words and uncertainties due to scientific controversies and contradictions as major sources of uncertainty. We justified the profound role of conflicting, contradictory, and inconsistent information in the course of scientific inquiry from philosophical and sociological perspectives and demonstrated the complexity of capturing the epistemic status of scientific propositions through a large-scale repository of semantic predications – the Semantic MEDLINE. We proposed a scalable and adaptive method to identify cue words for the study of the types of uncertainties in light of our conceptual framework. We manually compiled a seed list of uncertainty cue words and then used two Word2Vec models, based on Google News and PubMed, to generate an expanded list of candidate words. The candidate words are validated by two judges to accept and reject them as new cue words of uncertainty. The three types of words, namely, the seeds, accepted, and rejected cue words, are visualized and grouped together to form clusters of semantically similar words.
Our study aims to underline the significance of the study of uncertainties expressed or implied in scientific literature and how uncertainties evolve as new research published. The proposed conceptual framework attempts to build on existing research in fields such as computational linguistics, machine learning, scientometrics, and the study of scientific knowledge and focus on the profound role of uncertainty in scientific inquiry. In particular, we intend to draw attention towards the study of uncertainties due to inconsistencies, controversies, and contradictions because such uncertainties tend to have a greater degree of impact on scientific knowledge beyond individual scientific claims.
The proposed scalable and adaptive method for identifying uncertainty cues is only one step towards the development of an integrative methodology to study uncertainties with a specific focus on the tension between alternative theories and competing paradigms. Computational linguistic studies have contributed a rich set of resources and tools such as BioScope (Vincze et al., 2008), BioCause (Mihăilă et al., 2013), the CoNLL 2010 Shared Task (Farkas et al., 2010), meta-knowledge (Thompson et al., 2011). Machine learning tools such as the Word2Vec models we used in this study provide new opportunities for us to explore new approaches.

The current study has limitations and we plan to continue to refine the methodology in this area. For example, we used two judges to evaluate the expanded cue words. A larger number of judges with more extensive training would be an option. Furthermore, we plan to make use of the gold standards of the variety of uncertainties annotated in the computational linguistic studies, notably Vincze et al. (2008), Farkass et al. (2010), Szavas et al. (2012), Thompson et al. (2011), to name a few, and construct an annotated corpus with a focus on the contradiction-induced uncertainties for facilitate future research. The Word2Vec models can be further improved, for example, by constructing Word2Vec models with scientific publications from multiple disciplines. Currently, the PubMed model is biased towards biomedical sciences and the Google News is in a genre that may not be fully representative of scientific publications in general. Uncertainty cues in this study are limited to single words. Further studies should consider more complex expressions and discontinued expressions used by scientists in their publications. The research of scientific uncertain from computational and machine learning perspectives is highly complex and challenging. At the same time, it is also potentially highly rewarding.

Our conceptual framework is generic and adaptive to accommodate methodologies tailored to specific disciplines. As we have shown with a simple 5-word query, different disciplines are likely to have different content-specific uncertainties as well as other factors such as writing styles and disciplinary cultures. Investigating the dynamics of uncertainties in a diverse range of disciplines may lead to useful insights in the development of science. Our approach is a holist perspective in that we are concerned with the truthfulness of propositions across scientific publications as well as concerning the uncertainty of individual propositions and claims. The holistic perspective emphasizes the role of a broad context and guides us towards issues concerning consistencies and consensus and, more importantly, the concrete and complex course to reach such status.

Our conceptual framework broadens the scope of the types of uncertainties that can be consistently studied through integrations of computational linguistic approaches and the study of scientific knowledge. In particular, the focus on uncertainties due to controversial and contradictory information is a distinct extension of research that has focuses on hedging and linguistic markers that are loosely coupled with the underlying scientific knowledge.

In conclusion, identifying and reconciling conflicting observations and contradictory information is central to the advance of science. The level of uncertainties associated with the process is expected to decrease in general. On the other hand, we emphasize the complexity of this research topic because we are dealing with scientific knowledge, which involves the most complex form of abstraction, argumentation, and articulation. Studying the role of uncertainties in the development of scientific knowledge may offer a fruitful and more focused way to pursue scientific knowledge. The results of the study, a seed list and an expanded list of uncertainty cue words, can be used to identify sentences that address propositions with uncertainties and to identify disciplines or fields of research that are particularly rich in documents with explicit uncertainty cues.

We should also make it clear that our method is not intended to identify all the possible uncertainties in scientific publications. On the contrary, our goal is to make theoretical and practical contributions so that more research along these lines can advance the start of the art in understanding and tracking the development of scientific knowledge. In terms of the theoretical contribution, we introduce the conceptual framework that can be extended by adding new types of uncertainties. In terms of the practical contribution, we contribute the method and resultant uncertainty cues to the relevant research community. The patterns observed in this study are merely the tip of the iceberg. We contribute these lists to the research community as shared community resources for studying uncertainties in scientific knowledge. Ultimately, the key to reduce the types of uncertainties in scientific knowledge is the key to increase the productivity of scientific activities and the quality of scientific inquiries because we will be able to pinpoint the problem we need to deal with more efficiently.

## Acknowledgements

This work was supported by the Science of Science and Innovation Policy (SciSIP) Program of the National Science Foundation (#1633286). This work was also supported by the Bio-Synergy Research Project (NRF-2013M3A9C4078138) of the Ministry of Science, ICT and Future Planning through the National Research Foundation.